\documentclass[aps,prb,showpacs,raggedbottom,nobalancelastpage,amssymb,superscriptaddress]{revtex4}
\usepackage{amsmath}
\usepackage{amsfonts}
\usepackage{graphicx}
\usepackage{appendix}
\usepackage{dsfont}
\usepackage{amssymb}
\usepackage{bm}
\usepackage{color}
\usepackage{soul}
\usepackage{natbib}
\usepackage{nccmath}
\usepackage{array}
\usepackage{multirow}
\usepackage{wasysym}

\newcommand{\beq}{\begin{equation}}
\newcommand{\eneq}{\end{equation}}



\begin{document}

\title{From Kondo effect to weak-link regime in quantum spin-1/2 spin chains}

\author{Domenico Giuliano}
\affiliation{ Dipartimento di Fisica, Universit\`a della Calabria Arcavacata di 
Rende I-87036, Cosenza, Italy}
\affiliation{ I.N.F.N., Gruppo collegato di Cosenza, Arcavacata di Rende I-87036, Cosenza, Italy }

\author{Davide Rossini}
\affiliation{Dipartimento di Fisica dell'Universit\`a di Pisa, Largo Pontecorvo 3, I-56127, Pisa, Italy}
\affiliation{I.N.F.N., Sezione di Pisa, Largo Pontecorvo 3, I-56127, Pisa, Italy}

\author{Andrea Trombettoni}
\affiliation{ CNR-IOM DEMOCRITOS Simulation Center, Via Bonomea 265, I-34136 Trieste, Italy}
\affiliation{Scuola Internazionale di Studi Avanzati (SISSA), Via Bonomea 265, I-34136 Trieste, Italy }
\affiliation{ I.N.F.N., Sezione di Trieste, Via Bonomea 265, I-34136 Trieste, Italy }

\date{\today}

\begin{abstract} 
  We analyze the crossover from Kondo to weak-link regime by means of a model of tunable bond impurities
  in the middle of a spin-1/2 XXZ Heisenberg chain. We study the Kondo screening cloud 
  and estimate the Kondo length by combining perturbative renormalization group approach 
  with the exact numerical calculation of the integrated real-space spin-spin correlation functions. 
  We show that, when the spin impurity is symmetrically coupled to the two parts of the chain
  with realistic values of the Kondo coupling strengths and spin-parity symmetry is preserved, 
  the Kondo length takes values within the reach of nowadays experimental technology in ultracold-atom setups. 
  In the case of non-symmetric Kondo couplings and/or spin parity broken by a nonzero magnetic field applied
  to the impurity, we discuss how Kondo screening redistributes among the chain as a function 
  of the asymmetry in the couplings and map out the shrinking of the Kondo length when the magnetic field
  induces a crossover from Kondo impurity  to weak-link physics. 
\end{abstract}

\maketitle

\section{Introduction}
\label{intro}

The Kondo effect has been first seen in conducting metals containing magnetic impurities, 
such as Co atoms; it consists in an impurity-triggered, low-temperature increase 
in the metal resistance\cite{kondo64,hewson,kouwenhoven01}. Physically,  
Kondo effect is the result of nonperturbative spin-flip processes involving 
the spin of a magnetic impurity and of the itinerant conduction electrons 
in the metal, which  results in the formation, for vanishing temperature, 
of a strongly correlated Kondo state between the impurity and the conduction electrons 
\cite{kondo64,hewson}. In the Kondo  state, spins cooperate to dynamically screen the magnetic moment of 
the impurity \cite{kondo64,hewson,noz_1}. The specific properties of the correlated state 
depend on, {\it e.g.}, the number of independent ``spinful channels'' of conduction 
electrons participating to the screening versus the total spin of the magnetic
impurity. Denoting the latter by $s$, when the 
number of independent screening channels $k$ is equal to $2s$, in the Kondo state
the impurity spin is perfectly screened, which makes the screened impurity  
act as a localized scatterer with well-defined single-particle phase shift at the Fermi level.  
This corresponds to the onset of  Nozi\`eres Fermi-liquid state \cite{noz_1,noz_2}; at  
variance, when $k > 2s$, the Kondo state 
is characterized by impurity ``overscreening", which determines its peculiar, non Fermi-liquid 
properties \cite{ludaff_1,affleck_1}. 

Over the last decades, the Kondo effect emerged as a 
paradigm in the study of strongly correlated electronic states, providing 
an arena where to test many-body techniques, both 
analytical and numerical \cite{bulla}. Also, the realization of a Kondo interaction
involving Majorana fermion modes arising at the endpoints of one-dimensional (1D) topological 
superconductors has paved the way to a novel, peculiar form of 
``topological'' Kondo effect, sharing many common features with the overscreened 
multichannel Kondo effect \cite{topo_1,topo_2,topo_3,topo_4}.
Besides its fundamental physics aspects, the Kondo effect has attracted a renewed 
theoretical as well as experimental interest, since it has been possible to realize it with controlled 
parameters in quantum dots with either metallic \cite{alivisatos96,kouwenhoven98,gg98_1,gg98_2}, 
or superconducting leads \cite{avish,choi,gbab}.
This led to the possibility of using the Kondo effect to design quantum circuits 
with peculiar conduction properties, such as a conductance reaching the maximum 
value allowed by quantum mechanics for a single conduction channel \cite{kouwenhoven01}.
 
Formally, the Kondo effect is determined by a renormalization group (RG) crossover between 
quantum impurity ultraviolet and infrared fixed points (corresponding to the Kondo state). 
Typically, for a spin-1/2 impurity, near the ultraviolet fixed point (high energy), the coupling between 
the quantum impurity and the spin of conduction electrons is weak, thus merely providing 
a perturbative correction to the decoupled dynamics of the two of them. At variance, 
near the infrared fixed point (low energy), the conduction electrons in the Fermi sea 
adjust themselves to screen  the impurity spin into a localized spin singlet. Regarding 
the relevant energy window for the process,   the impurity spin screening requires a cooperative effect of electrons with energies
all the way down to $k_B T_K$, with $k_B$ being the Boltzmann constant 
(which we set to $1$ henceforth) and $T_K$ the Kondo temperature, that is,   
a temperature scale invariant under RG trajectories and 
dynamically generated by the Kondo dynamics \cite{hewson}. At energies $\sim T_K$ 
a crossover takes place, between the perturbative dynamics of the impurity  spin weakly coupled 
to itinerant electrons and the nonperturbative onset of the Kondo state. 
The possibility of using the Fermi velocity $v_f$ to trade an energy scale $E_*$ for 
a length scale $L_* \sim v_f / E_*$ led to the proposal that the crossover might be 
observed in real space, as well \cite{wilson,noz_1}. Switching from 
an energy to a length reference scale implies that dynamical impurity spin screening has 
now to be thought of as a real-space phenomenon, with the net effect of substituting, as 
a reference scale for screening,  the temperature with the distance from the impurity $x$. 
In other words, a physical quantity depending on 
the distance $x$ from the impurity is expected (on moving away from the impurity) 
to exhibit a crossover from a perturbative behavior controlled by the ultraviolet fixed point at small $x$ to 
a nonperturbative behavior controlled by the Kondo state at large $x$, the crossover taking place 
at a scale $\xi_K \sim v_f / T_K$. Such value $\xi_K$ can accordingly be regarded as the size of 
the electronic cloud screening the impurity spin: for this reason, it is typically referred to as the 
{\it Kondo screening length}. 

The presence of a Kondo cloud (KC) is well-grounded on the theory side and 
it has also been recently proposed that an analogous phenomenon
takes place at a Majorana mode coupled to a 1D quantum wire \cite{giuaf_3}. 
However, any attempt to experimentally detect it at magnetic impurities in metals has so far failed. 
There is a number of possible reasons for that: first of all, from typical values of 
$T_K$ in metals, $\xi_K$ is estimated to be of the order of thousands metallic lattice spacings, 
which makes spin correlations between the impurity and the itinerant electrons in 
practice not detectable as $x \sim \xi_K$. In addition,   in real metals one 
typically recovers a finite density of magnetic impurities. Thus,  the large value of $\xi_K$ very likely 
implies interference effects between clouds relative to different impurities. Also, 
the simple models one uses to perform the calculations may be too simplified lacking, for instance, 
effects of electronic interactions, etc. (for a review about the Kondo screening cloud, 
see Ref.~[\onlinecite{Affleck09}] and references therein). For these reasons, the 
quest for the Kondo cloud has recently moved to realizations of 
the Kondo effect in systems different from metals, such as quantum spin chains.

In fact, it is by now well established that 
the Kondo effect can be achieved in magnetic impurities coupled to 
antiferromagnetic spin chains with a gapless spin excitation spectrum, the so called {\it spin-Kondo effect}
\cite{eggert92,furusaki98,sorensen}.
Indeed, the Kondo effect is merely due to spin dynamics \cite{sorensen,Affleck09} and, 
because of spin fractionalization \cite{tak,hal}, 
a quantum antiferromagnetic spin chain can be regarded as a sea of weakly interacting 
collective spin-1/2 excitations with a gapless spectrum, dubbed spinons \cite{hal,gbl}, 
which eventually cooperate to dynamically screen the spin of the magnetic 
impurity in the chain. Besides its interest {\it per se}, the spin-Kondo effect 
also  provides an effective description of Kondo-like dynamics 
in a number of different physical systems that have been shown 
to be effectively described as a (possibly inhomogeneous) XXZ spin chain,
such as the Bose-Hubbard model realized by loading cold atoms onto an 
optical lattice \cite{grst_13,gst}, as well as networks made joining together 
1D arrays of quantum spins or of 
quantum Josephson junctions \cite{giuliano07,giulianoepl,giuliano09,crampe,gst_1}.
Also, studying Kondo effect in spin chains allows for investigating various 
aspects of the problem relevant to quantum information  such as, 
for instance,  entanglement witnesses and negativity \cite{pask_1,pask_2}.
To date, different realizations of Kondo effect in spin chains have 
been considered  in the case in which an isolated magnetic impurity is side-coupled to a single uniform XXZ-chain
(single-channel Kondo-spin effect) \cite{eggert92}, 
to a frustrated $J_1-J_2$ antiferromagnetic spin chain \cite{sorensen} and 
to a ``bulk'' spin in an XXZ spin chain \cite{furusaki98}.

In this paper, we consider a magnetic impurity realized in the middle of the chain by weakening  two consecutive 
bonds in an otherwise uniform XXZ chain with open boundaries. Notice that, 
to have the spin-chain Kondo effect, one needs to have a {\it single} 
bond impurity ({\it i.e.}, an altered and decreased 
coupling between two neighboring sites) on the edge -- or {\it two} bond impurities in the middle ({\it i.e.}, 
in the bulk) of the chain, as we are going to discuss. Remarkably, 
on the experimental side, the recent solid-state construction of 
an XXZ-spin chain using Co-atoms deposited onto a 
CuN$_2$/Cu(100)-substrate \cite{otte_2}  paves the way to a 
realistic experimental realization of the system we discuss. On the theoretical side, with respect
to the previous systems listed above, our proposed system presents a number of features that motivate
an extensive treatment of the corresponding realization of Kondo effect.
First of all, we consider a magnetic impurity separately coupled to two 
independent screening channels, that is, the spin-chain version of the two-channel Kondo effect 
\cite{eggert92,Affleck94,pikulin}; this allows us, by tuning the couplings to the two channels, 
to move from two-channel to one-channel spin-Kondo effect, and back. As a result, 
it  enables us to study, for the first time, how screening sets in and 
is distributed among the channels in a multichannel realization of Kondo effect. 
Moreover, we show that acting upon an applied magnetic field at 
the impurity, allows for switching from a Kondo system to a 
simple weak-link between two otherwise homogeneous spin chains, thus 
allowing for mapping out the effects on $\xi_K$ when crossing over 
between the two regimes. Specifically, we combine the analytical approach based on
the perturbative RG equations, which we derive in 
the case of a nonzero applied magnetic field at the impurity, with 
a density-matrix renormalization group (DMRG) based 
numerical derivation of a suitable integrated  
real-space spin-spin correlation between the magnetic impurity and the spins
of the chains. In Ref.~[\onlinecite{baraf}] 
a similar quantity was originally proposed as a mean to directly provide the 
Kondo screening cloud through its scaling behavior in real space. Here, we construct 
a version of the integrated correlation function that is suitable for 
a Kondo impurity in an XXZ quantum spin chain. This is an adapted version 
of the function used to extract, from numerical data, 
the Kondo screening length at an Anderson impurity lying at the endpoint of a 1D
lattice electronic system \cite{sholl}.

The combination of the analytical and numerical methods allows us to properly choose the 
ultraviolet cutoff entering the solution of the RG equations in 
the various cases. Doing so, we recover an excellent consistency between 
the analytical and the numerical results, which allowed us to derive analytical 
scaling formulas for the integrated correlation functions in all the cases in
which a pertinent version of perturbation theory is expected to apply.   
Performing a  scaling analysis of the integrated correlation function,
we generalize the formalism of Refs.~[\onlinecite{baraf,sholl}] to Kondo effect 
in a quantum spin chain. Due to the remarkable mapping between an XXZ spin chain 
and a 1D single-component Luttinger liquid, which also describes 
interacting spinless electrons in one spatial dimension \cite{eggert92}, by 
the same token we also show how to generalize the method of Refs.~[\onlinecite{baraf,sholl}] 
to Kondo effect with interacting electronic leads. Within our technique, we 
prove that, at physically reasonable values of the Kondo couplings $\xi_K$ in a 
spin chain ranges from a few tens, to about $100$ times the lattice spacing. 
We also provide for the first time, to the best of our knowledge, 
a detailed qualitative and quantitative description of the 
behavior of the screening cloud in two-channel spin-Kondo effect, as well as 
of the shrinking of the cloud when an applied magnetic field at the impurity 
makes the system switch from a Kondo impurity to a weak-link between 
homogeneous spin chains.

About focusing our discussion on Kondo effect in spin chains it is, finally, 
worth stressing that magnetic impurities in spin chains mimic the Kondo effect 
in the precise sense that the low-energy model describing their dynamics is the same as the conventional Kondo model \cite{sorensen}. 
So, in light of this mapping, most of the results we are going to discuss have a precise counterpart in other realizations
of the Kondo effect, and similarly the RG treatment presented here can be performed in other Kondo contexts. Having stated 
so, it is worth stressing that studying Kondo effect in spin chains offers two important advantages: 

{\it i)} It allows for computing correlation functions using DMRG, thus admitting a convenient exact numerical 
benchmark for the analytical results; 

{\it ii)} It makes it possible to propose an experimental setup for realizing the crossover between 
different impurity regimes, such as a Kondo impurity and a single weak link, which we discuss 
in the paper, by  physically simulating  
various quantum spin chains with ultracold atoms in optical lattices (see, for instance Ref.~[\onlinecite{grst_13}] and references therein). 
In particular, the Bose-Hubbard model \cite{lewenstein} at half-filling can be mapped onto the XXZ spin chain 
\cite{matsubara,fazio95,grst_13} and values of the Kondo length of order $10-100$ lattice sites 
can be conceivably detected in ultracold atom experiments by extracting the Kondo length from correlation functions~\cite{gst}.

The paper is organized as follows:

\begin{itemize}
\item In section~\ref{moha} we introduce the model Hamiltonian we use throughout all the paper. We discuss 
  the physical meaning of the Hamiltonian parameters, introduce the running Kondo coupling strengths and 
  outline how they are used to estimate $\xi_K$.

\item In section~\ref{icf} we define the integrated spin correlation function and discuss how to use it to 
  estimate $\xi_K$. In particular, in Sec.\ref{scacol}, we discuss 
  the scaling collapse technique, while in the following section, Sec.\ref{klc}, we review the Kondo length 
  collapse method. Throughout all Sec.\ref{icf}, we limit ourselves to the case of symmetric Kondo 
  couplings and zero applied magnetic field at the impurity.

\item In Sec.\ref{nonsymmetric}, we generalize the results of Sec.\ref{icf} to the case of 
  non-symmetric Kondo couplings (Sec.\ref{asym}), as well as to the case of a nonzero magnetic 
  field applied to the impurity (Sec.\ref{magimp}).

\item In Sec.\ref{concl}, we summarize our results and discuss possible further developments of our work.

\item In the various appendices we discuss mathematical details, such as the 
  mapping between extended spin clusters in the chain and effective Kondo- or weak-link
  impurities, the spinless Luttinger liquid approach to the XXZ spin chain and its application to 
  derive the RG equations in the case of a Kondo impurity, as well 
  as of a weak-link between two chains.
\end{itemize}

\section{Model Hamiltonian}
\label{moha}

Our main reference  Hamiltonian ${\cal H}$ describes  a magnetic spin-1/2 impurity
${\bf S}_{\bf G}$ embedded within an otherwise uniform XXZ spin chain. We assume that 
the applied magnetic field along the chain is zero everywhere but at the impurity location, where 
it takes  a nonzero value $B$ in the $z$ direction. To avoid unnecessary 
computational complications, we assume that the whole chain consists of an odd number of sites, 
$2 \ell + 1$, with integer $\ell$, and ${\bf S}_{\bf G}$ sitting in the middle of the 
chain. At each site lies a spin-1/2 quantum spin degree of freedom: 
we denote with ${\bf S}_{j , L}$ and 
with ${\bf S}_{j , R}$ the corresponding vector operators sitting at site $j$, measured from the impurity 
location (which accordingly we set at $j=0$), on respectively the left-hand and the right-hand side of the chain. 
As a result,  ${\cal H}$ takes the form ${\cal H} = \sum_{ X = L , R} H_X + H_K$, with 
\begin{eqnarray}
 H_X &=& J \sum_{ j = 1}^{\ell - 1 } \{ S_{ j , X }^+ S_{j + 1 , X}^- + S_{j , X}^- S_{j + 1 , X}^+ + \Delta S_{ j , X}^z 
 S_{j + 1 , X}^z \} \nonumber \\
 H_K &=& \{ J^{'}_L  S_{ 1 , L}^+ + J^{'}_R  S_{ 1 , R}^+ \} S_{\bf G}^- +  \{ J^{'}_L  S_{ 1 , L}^-  + J^{'}_R  S_{ 1 , R}^- \} S_{\bf G}^+ 
 +  \{ J^{'}_{z ,L}   S_{ 1 , L}^z + J^{'}_{z ,R} S_{ 1 , R}^z \} S_{\bf G}^z  + B S_{\bf G}^z \; . 
 \label{modham.1}
\end{eqnarray}
The spin operators ${\bf S}_{j , L (R)}$ in Eq.~(\ref{modham.1})  satisfy the algebra 
\beq
S_{j , X}^a \, S_{j' , X'}^b = \delta_{j , j'} \, \delta_{X , X'} \left\{ \frac{ \delta^{a,b} }{4} + \frac{i}{2}  \epsilon^{abc} S_{j , X}^c \right\} \; , 
\label{mh.1}
\eneq
\noindent 
with $X , X' = L , R$, $a , b, c  = x , y , z$ and $\epsilon^{abc}$ being the fully antisymmetric tensor. 
$J$ is the (antiferromagnetic) exchange strength, providing an over-all 
energy scale of the system Hamiltonian. The anisotropy $\Delta$ is the ratio between the 
exchange strengths in the $z$  and in the $xy$ directions in spin space. In order to recover 
spin-Kondo effect, one has to avoid the onset of either antiferromagnetically, or ferromagnetically, 
ordered phases in the chain, which requires 
(as we do throughout the whole paper) assuming $-1 \leq \Delta \leq 1$ \cite{luther}. 
${\bf S}_{\bf G}$ is coupled to the rest of the chain via the boundary, transverse and longitudinal ``Kondo'' couplings,
respectively, given by $J_{L}^{'},J_{R}^{'}$ and by $J_{z , L}^{'},J_{z , R}^{'}$. To achieve Kondo physics,   
 $J_{L ( R)}^{'}$ and  $J_{z , L(R)}^{'}$ must respectively be smaller than $J$ 
and than $\Delta J$ (this is necessary to ``leave out'' some room for the onset of the perturbative Kondo regime, 
which is, in turn, necessary in order to define the Kondo screening length \cite{barzykin,Affleck09}).
Again, to avoid unnecessary computational complications, as for the ``bulk'' parameters of the chain, we 
set $J_{z , L (R)}^{'} =  \Delta J_{L ( R)}^{'}$. 
  
The model in Eq.~(\ref{modham.1}) corresponds to a two-channel Kondo-spin 
Hamiltonian \cite{eggert92,Affleck94,furusaki98,sorensen,pikulin}, in which the impurity spin ${\bf S}_{\bf G}$
is independently coupled to a two spin-1/2 ``spinon baths'', at the two sides 
of $ {\bf S}_{\bf G}$ \cite{hald_1}. Accordingly, ${\cal H}$ allows to study how the 
Kondo screening is affected by {\it e.g.} an asymmetry in the Kondo couplings to the 
two channels, as well as by a nonzero $B$ applied to the impurity site and, eventually, to map 
out the crossover from Kondo- to weak-link physics at the impurity. 
 At this stage, it is worth pointing out one of the key differences between spin-1/2 multichannel Kondo effects 
in systems of itinerant electrons and in spin chains. In our spin-chain model Hamiltonian, when ${\bf S}_{\bf G}$ is symmetrically coupled 
  to the spin densities from the XXZ-chains, each chain works as an independent spin screening channel. Yet, 
differently from what happens in electronic Kondo effect, where the overscreening (that is, $k>2s$) leads to the onset of a nontrivial
finite coupling fixed point, in spin chain it only trades into a symmetric healing of the spin chain, with the Kondo fixed point simply 
corresponding to an effectively uniform chain \cite{eggert92,Affleck94,pikulin}.

The Kondo effect is triggered by the fact that, depending on the value of $\Delta$, $H_K$ either realizes 
a relevant or a marginally relevant boundary perturbation, which eventually leads to the emergence of the
dynamically generated length scale $\xi_K$. Resorting to the spinless Luttinger liquid (SLL) representation of the homogeneous chains at 
the sides of ${\bf S}_{\bf G}$ (the ``leads''), the key parameter determining the behavior of the 
boundary interactions is the Luttinger parameter $g$, which is related to $\Delta$ via 
(see Appendix~\ref{sll} for details)
\beq
g = \frac{\pi}{2 [ \pi - {\rm arccos} (\Delta  )]}  \, . 
\label{mh.01}
\eneq
\noindent 
Specifically, $H_K$ is marginally relevant when $g=1/2$ (corresponding to $\Delta = 1$, that is, to 
the $SU(2)$-isotropic XXX spin chain), while it becomes relevant as soon as  
$g > 1/2$ \cite{eggert92,furusaki98,sorensen}. As we are eventually interested in regarding 
the XXZ-Hamiltonian as an effective description of cold-atom bosonic lattices \cite{gst}, throughout all 
the paper we will assume $g > 1/2$. Incidentally, this also enables us to rely on the Abelian bosonization 
approach, which corresponds to the SLL-formalism of appendix~\ref{sll}, rather than resorting to the 
more complex and sophisticated non-Abelian bosonization scheme, which is more suitable to provide 
a field-theoretical description of the isotropic XXX chain \cite{eggert92}. 

The relevance of $H_K$ is encoded in the RG flow of the boundary 
couplings associated to $H_K$  (for an extensive discussion about this point, see, for instance, Ref.~[\onlinecite{hewson}]). 
In the specific case of Eq.~(\ref{modham.1}), the running couplings are given by 
$G_{L (R)}  ( \ell )  =  \frac{1}{2} \left( \ell / \ell_0 \right)^{1 - \frac{1}{2g} } \frac{J'_{ L ( R ) }}{J}$
and $G_{z , L (R)}  ( \ell ) = \frac{1}{2} \: \frac{J'_{z ,  L ( R ) }}{J}$, where 
$\ell_0$ is the short-distance cutoff. In appendix~\ref{reng} we 
discuss the derivation of the RG equations. In particular, we stress the emergence of 
  $\xi_K$ as the scale  at which the running couplings enter 
the nonperturbative regime. Note that, since we are eventually interested in 
deriving the expression for $\xi_K$, throughout our paper 
we use the system size $\ell$, rather than the temperature $T$,  as the scale parameter 
triggering the RG flow of the running coupling strengths. This basically corresponds to setting $T=0$ and keeping 
$\ell$ finite or, more generally, to assuming that $k_B T \ll v_f / \ell$.  
The physical interpretation of $\xi_K$ as the size of the Kondo cloud stems from  Nozi\`eres picture 
of the Kondo screening cloud, in which spins surrounding ${\bf S}_{\bf G}$ over a distance $\sim \xi_K$ 
cooperate to screen the magnetic impurity into the extended Kondo singlet \cite{noz_1,noz_2}. 
It is worth now considering the physical picture of the fixed point toward which 
the system is attracted, when crossing over to the strongly coupled regime (that is, as soon as $\ell \sim \xi_K$). 
To begin with, let us focus onto the $B=0$ and $L-R$-symmetric case: $J_L^{'} = J_R^{'}$ and 
$J_{z , L}^{'} = J_{z , R}^{'}$. In this case, one expects the onset of   
a two-channel Kondo regime, in which the ${\bf S}_{\bf G}$ is 
equally screened by spins at both sides of the impurity, as soon as 
$\ell \geq \xi_K$. The corresponding Kondo fixed point can be 
easily recovered as being equivalent to an effectively uniform chain, 
as all the possible boundary perturbations preserving the $L-R$-symmetry become 
an irrelevant perturbation at such a fixed point \cite{eggert92,hikifu_1}. At variance, 
different fixed points are realized either when the $L-R$-symmetry 
is broken, or there is a nonzero magnetic field $B$ applied to ${\bf S}_{\bf G}$ (or both). 
The former case is realized when, for instance,  
$J_L^{'} <  J_R^{'}$ and $J_{z , L}^{'} < J_{z , R}^{'}$. In this 
case, one may attempt to define a left-hand and a right-hand screening length, 
respectively referred to as $\xi_{K , L}$ and as $\xi_{K , R}$. From the 
explicit formulas in Eqs.~(\ref{mh.10}, \ref{mh.12}, \ref{mh.14}) of appendix~\ref{reng}, 
one therefore expects that $\xi_{K,L} > \xi_{K , R}$. This implies that, on 
equally increasing $\ell$ on both sides of the impurity spin, the condition $\ell \sim \xi_{K , R} $ is 
met first. As $\ell \sim \xi_{K , R}$, ``healing'' of the weak-link between ${\bf S}_G$ and ${\bf S}_{1 , R}$ is
complete, and one may accordingly regard the whole system as an $\ell + 1$-site uniform chain, made out of 
the $\ell$ sites hosting the ${\bf S}_{j , R}$ spins plus ${\bf S}_{\bf G}$, coupled at its endpoint to an 
$\ell$-site chain -- made out of the $\ell$ sites hosting the ${\bf S}_{j , L}$ spins -- via the ``residual'' 
weak-link boundary Hamiltonian $H_{\rm W}$ given by  
\beq
H_{\rm W} = \bar{J}' \{ S_{\bf G}^+ S_{1 , L}^- + S_{\bf G}^- S_{1 , L}^+ \} 
+ \bar{J}'_z S_{\bf G}^z S_{1 , L}^z \, , 
\label{mh.15}
\eneq
where $\bar{J}' , \bar{J}_z^{'}$ are defined from the running couplings $G_L ( \ell ) , G_{z , L} ( \ell )$
at $\ell \sim \xi_{K , R}$. 

$H_{\rm W}$ in Eq.~(\ref{mh.15}) is the prototypical ``weak-link'' boundary 
Hamiltonian we review in appendix~\ref{wlink} \cite{giuliano05,eggert99}. To address its behavior on 
further rescaling $\ell$, one defines the novel running dimensionless couplings  
$\Gamma ( \ell ) =\left( \ell / \xi_{K , R} \right)^{1 - 1/g} \: \bar{J}'$ and 
$\Gamma_z ( \ell ) = \bar{J}_z^{'}$. For $g<1$, standard RG  approach
implies that $H_{\rm W}$  corresponds to an irrelevant boundary interaction \cite{kanelett,kane92,rommer,giuliano05}.
This on one hand implies that no additional length scales associated to screening are dynamically 
generated along the RG flow of $\Gamma ( \ell )$ and of  $\Gamma_z ( \ell )$, on the other hand that
any physically relevant quantity, such as the real space correlations between the $_L$ and the $_R$ spins,
can be reliably computed in a perturbative expansion in $H_{\rm W}$. At variance,  
in the case $g>1$ (which corresponds to $\Delta < 0$), $H_{\rm W}$ does become a relevant operator. 
However, this only quantitatively affects 
the final result, in that the weak-link couplings now become effectively dependent 
on the scale. Again, the ``healing'' of the chain \cite{glark,giuliano05} sets in without 
the onset of any Kondo cloud and, again, no scaling is expected to be seen in 
the correlations. More generically, for 
any value of $g$ we expect weak-link physics to apply, with no additional length scales 
being dynamically generated. This  can be discussed in close analogy to Kondo effect in metals. 
There,  $\xi_K$ emerges at the crossover to the nonperturbative regime inside the
Kondo cloud. Outside the Kondo cloud, 
the impurity spin is screened and what one sees is the residual interaction 
corresponding to Noz\`ieres Fermi liquid \cite{noz_1,noz_2}, with  no additional scales 
dynamically generated. 
So, we may pictorially state that our weak-link is, in a sense, the analog, for 
the spin chain, of what Nozi\`eres Fermi liquid is for Kondo effect in metals.
 
A similar physical scenario is realized in the case of a nonzero applied 
$B$ which, as we discuss in appendix~\ref{wlink}, again yields an effective weak-link 
Hamiltonian at the impurity. To get a qualitative understanding, we 
note that $B \neq 0$ induces an additional length scale $\xi_B \propto J / B$. At small values of $B$, one 
typically has $\xi_K \ll \xi_B$, which implies that Kondo effect is not substantially affected, as long 
as $B / J \ll 1$. In fact, as we discuss below, a finite $B$ merely provides a slight renormalization of 
$\xi_K$ which, in a sense, is analogous to what happens to electronic Kondo effect when the single-electron 
spectrum has a gap $\bar{E}$ at the Fermi level, but the Kondo temperature is still much larger than
$\bar{E}$ \cite{avish,choi,gbab,gct}.
At variance, as we discuss in the following, a substantial suppression of the Kondo effect and 
a switch to weak-link physics, with a corresponding collapse of the screening length, is 
expected for $\xi_B \leq \xi_K$ \cite{costi,sholl}.  

To conclude this section, it is worth stressing the  important point about ${\cal H}$, 
addressed in detail in appendix~\ref{mappings}, that, besides describing a single spin in a generically nonzero 
magnetic field weakly coupled to two uniform spin chains, it can be regarded as an effective description 
of a generic few-spin spin cluster (an ``extended region'') in the middle of an otherwise uniform chain, weakly 
coupled to the rest of the chains at its endpoints. An extended region is closer to what one 
expects to realize in a bosonic cold-atom lattice \cite{gst}. In such systems, one can in general 
simulate quantum spin models by loading the quantum gas(es) on optical lattices \cite{lewenstein}. To map 
the resulting Bose-Hubbard Hamiltonian on an XXZ model, there are several possible strategies; we refer 
in particular to the one proposed in Ref.~[\onlinecite{grst_13}], where the 1D Bose-Hubbard Hamiltonian 
at half-filling is mapped on an effective XXZ model, and their correlation functions compared finding 
remarkable agreement also for interaction strength relatively small. In this physical setup 
the antiferromagnetic couplings $J$'s on the links are proportional to the tunneling rates 
for the bosons hopping from one well to its nearest neighbor. So one can locally alter the tunnelings by 
adding one or two repulsive potentials 
via laser beam (see Fig.1 of  Ref.~[\onlinecite{grst_13}]). Denoting by $\sigma$ 
the spatial $1/e^2$ beam waists of the lasers one has different situations: 
denoting by $d$ the lattice spacing, if $\sigma \lesssim d$ and one has just one laser centered 
on the maximum of the energy barrier between two minima ({\it i.e.}, two sites), 
then one is practically altering only one coupling $J$. When one has 
two lasers,  with intensities denoted by, say, $V_L$ and $V_R$, and again $\sigma \lesssim d$ 
then one is altering two couplings, and, depending on  the precision with which one is centering the lasers,  one can 
have equal ($J_{L}^{'} \sim J_{R}^{'}$) 
or different couplings ($J_{L}^{'} \neq J_{R}^{'}$). When $\sigma \gtrsim d$ 
then one is unavoidably altering several links, leading to an approximately Gaussian deviation 
of the couplings from the left and right bulk coupling $J$ extending roughly on $\sigma / d$ sites. 
We refer to Ref.~[\onlinecite{grst_13}] for details, but typically $\sigma \gtrsim 2\mu m$, and $d$ is order of 
$0.5-1\mu m$, even though by having tunable barriers one can have appreciable tunnelings also for $2-3 \mu m$ 
or larger \cite{dw}.  In summary, for $\sigma \lesssim d$, 
one approximately has that for: {\em i)} $V_L=0$ and $V_R \neq 0$, 
all couplings are equal (to $J$) and one link is altered ($J_{R}^{'}$); 
{\em ii)} $V_L=V_R \neq 0$, all couplings are equal (to $J$) but the two central ($J_{L}^{'}=J_{R}^{'}$); {\em iii)} fixed $V_{R}\neq 0$ and varying $V_{L}$, 
interpolates between the single non-magnetic altered bond providing 
a non-magnetic weak-link ($V_{L}=0$) and the case in which the chain is cut 
in two and there is a single altered bond in the right-half of the chain 
($V_{L} \gg V_{R}$), behaving at variance as a magnetic impurities and giving 
rise to the one-channel Kondo effect \cite{sorensen}.

Therefore, we see that, in realizations with ultracold atoms one would generically have an extended region, 
even though altering few tunneling terms is conceivable. Yet, ${\cal H}$ is expected to 
be able to catch the relevant physical behavior, as well, provided its parameters are carefully set 
by, {\it e.g.}, following the route we illustrate in appendix~\ref{mappings} in some specific paradigmatic cases.
Eventually, this makes ${\cal H}$ in Eq.~(\ref{modham.1}) to be worth studied as a paradigmatic 
effective description of an extended region in an otherwise uniform spin chain.

\section{Kondo screening length from integrated real space correlation: \newline the $B=0$ and $_L$-$_R$ symmetric limit}
\label{icf}

In this section we illustrate in detail how to construct and use the integrated real-space 
correlation function ${\bf \Sigma} [ x ]$ to probe the Kondo screening cloud in real 
space and to eventually extract the corresponding value of $\xi_K$. In order to do so, 
we refer to the so far firmly established scaling properties (with $\xi_K$) of the real-space correlations 
between ${\bf S}_{\bf G}$ and the screening spins from the leads, which  
have been put forward by making a combined use 
of perturbative RG methods, as well as of fully  numerical DMRG approach \cite{affleck_length,barzykin}.
As an extension of the results of Refs.~[\onlinecite{affleck_length,barzykin}], 
Barzykin and Affleck have proposed to look at the scaling properties of 
the integrated real-space correlation function  as a mean to directly map out the 
Kondo screening cloud in real space \cite{baraf}. Following their proposal, we 
introduce the function ${\bf \Sigma} [ x ]$  as 
an adapted version of the integrated correlation function originally proposed in Ref.~[\onlinecite{baraf}]
to discuss Kondo cloud at an isolated magnetic impurity in a metal, and later on adapted to 
an Anderson impurity lying at the endpoint of a 1D lattice electronic system \cite{sholl}.
In defining ${\bf \Sigma} [ x ]$, we  necessarily have to take into account that, even as $B = 0$, the 
easy-plane anisotropy of the XXZ-chain at $ | \Delta | < 1$ 
breaks the spin $SU(2)$-symmetry, leaving as a residual symmetry the
group $U ( 1 )$ associated to rotations around the $z$-axis in spin space. Following Ref.~[\onlinecite{sholl}], 
we therefore set  
\beq
{\bf \Sigma}  [ x ] = 1+  \sum_{i = 1}^{x} \sum_{X = L , R}  \left\{ \frac{\langle S_{\bf G}^z S_{i , X}^z \rangle - 
\langle S_{\bf G}^z \rangle \langle S_{i , X}^z \rangle }{
\langle (S_{\bf G}^z )^2 \rangle - ( \langle S_{\bf G}^z \rangle )^2 } \right\}  
\, . 
\label{icf.1}
\eneq
\noindent 
A full $SU(2)$-symmetric version of ${\bf \Sigma} [ x ]$ (which apparently does not apply to the system 
we consider here), would be given by  \cite{baraf,sholl}
\beq
{\bf \Sigma}^{SU(2)} [ x ] = 1 + \sum_{i = 1}^{x} \sum_{X = L , R}  \left\{ \frac{\langle {\bf S}_{\bf G}  \cdot  {\bf S}_{i , X} \rangle -
\langle {\bf S}_{\bf G} \rangle 
\cdot \langle {\bf S}_{i , X}  \rangle }{
\langle ({\bf S}_{\bf G} )^2 \rangle - ( \langle {\bf S}_{\bf G} \rangle )^2 } \right\}
\, . 
\label{icf.2}
\eneq
\noindent 
Due to the normalization we use in Eq.~(\ref{icf.1}), one has ${\bf \Sigma} [ x = 0 ] = 1$ while, 
since the state on which we compute spin correlations is an eigenstate, or a linear 
combination of eigenstates, of $S_T^z = S_{\bf G}^z + \sum_{j = 1}^\ell \{ S_{j , L}^z + S_{j , R}^z \}$, 
one recovers 
the second boundary condition ${\bf \Sigma} [ x = \ell ] = 0$ \cite{sholl}. When moving from the 
impurity location, ${\bf \Sigma} [ x ]$ is expected to show a net decreasing, due to the screening of ${\bf S}_{\bf G}$ 
by spins in the leads. In fact, this is the case though, for $0 < \Delta < 1$, the antiferromagnetic 
spin correlations make the decreasing to be not monotonic, but characterized by a staggering by one lattice step, 
with a net average decrease as $x$ increases \cite{sholl}. When $\ell \gg \xi_K$ one expects that finite-size effects are suppressed and, 
therefore, that, as long as $x < \xi_K$, 
${\bf \Sigma} [ x ]$ probes the inner part of the Kondo cloud. The farther one moves from the impurity (increasing 
$x$), the more one enters the nonperturbative regime, till one eventually recovers full Kondo screening, as soon as 
$x \sim \xi_K$. At variance, for $x > \xi_K$, ${\bf \Sigma} [ x ]$ probes the region outside of the Kondo cloud. 
This latter region corresponds to Nozi\`eres Fermi liquid theory for the Kondo fixed point, with a completely 
different expected behavior of the scaling properties of  ${\bf \Sigma} [ x ]$  \cite{noz_1,noz_2}.
Basically, one can state that the behavior of ${\bf \Sigma} [ x ]$ is described by: 

{\it i)}  the weakly coupled fixed point
(${\bf S}_{\bf G}$ weakly coupled to the chains) for $ x / \xi_K \ll 1$  
(note that this is profoundly different from the case of a boundary interaction effectively behaving as  a
single weak-link, in which one does not expect any particular dynamically generated emerging length scale to be 
associated with scaling  properties of the Kondo cloud  \cite{matvegla_1,matvegla_2,giu_nava,gst});

{\it ii)} the strongly coupled Kondo fixed point (uniform chain limit, corresponding to Nozi\`eres fixed point  
for electrons in a metal) for $ x / \xi_K \gg 1$ \cite{giuaf_3}. 

To better illustrate the application of our method, in this section we set $B = 0$ and focus on a system with 
symmetric boundary couplings $J_L' = J_R' = J'$ and $J_{z , L}^{'} = J_{z , R}^{'} = \Delta J'$. For $x \ll \xi_K$,
we estimate ${\bf \Sigma} [ x ]$ to leading  order in $J'$. Within SLL-framework of appendix~\ref{sll}, we obtain

\beq
{\bf \Sigma} [ x ]  \approx 1 + 
\sum_{j = 1}^x \:  \frac{   J_{z }^{'}}{2}  \left\{  \frac{g}{2   \pi u \ell} -  \frac{2^{1 - g} a g}{\ell} 
  \left[ \frac{ \sin \left( \pi j / \ell \right) }{1 - \cos \left( \pi j / \ell \right)  } \right] -  \frac{2 g a [ 1 + 2^{-g} \pi a ]
  (-1)^j j }{u \ell} \: 
  \left| \frac{2 \ell}{\pi} \sin \left( \frac{\pi j}{\ell} \right) \right|^{-g} \right\} 
\, . 
\label{wcf.00}
\eneq
\noindent 
To encode the perturbative RG results in Eq.~(\ref{wcf.00}), we 
follow the ``standard'' strategy \cite{hewson} of substituting the ``bare'' couplings $J_{z }^{'}$ with the 
running one, $G_{z} \left( \frac{x}{\xi_K} \right)$, obtained from Eqs.~(\ref{mhscle}) 
of appendix~\ref{reng}, in which the dependence on $\ell_0$ has been traded for a dependence on $\xi_K$ (see 
appendix~\ref{reng} for details), and $x$ is used as infrared cutoff, consistently with the 
fact that one has to integrate of a spin cluster of size $\sim  x$ \cite{baraf,gst}. For large $\ell$, we may trade 
the sum in Eq.~(\ref{wcf.00}) for an integral, getting 
\beq
{\bf \Sigma} [ x ]  \approx 1 + \frac{  G_{z} \left( x / \xi_K \right)}{2} \! \int_{\ell_0}^x \!\! d w \!
  \left\{  \frac{g}{2   \pi u \ell} -  \frac{2^{1 - g} a g}{\ell}
  \left[ \frac{ \sin \left( \pi w / \ell \right) }{1 - \cos \left( \pi w / \ell \right)  } \right] -  \frac{2 g a [ 1 + 2^{-g} \pi a ]
  w \cos ( \pi w )  }{u \ell} \, 
  \left| \frac{2 \ell}{\pi} \sin \left( \frac{\pi w}{\ell} \right) \right|^{-g} \right\} \, . 
\label{wcf.0a}
\eneq
\noindent 
Following the approach of Ref.~[\onlinecite{baraf}], from Eq.~(\ref{wcf.0a}), we infer a general scaling formula for 
${\bf \Sigma} [ x ]$ when $x / \xi_K < 1$, given by 
\beq
{\bf \Sigma}  [ x ] = \sum_{ d_b } \ell^{d_b } \xi_{d_b} \left[ \frac{\xi_K}{\ell} ; \frac{x}{\xi_K} \right] 
\, , 
\label{icf.3}
\eneq
\noindent 
with the sum taken over the scaling dimensions of the boundary operators entering 
the SLL representation for $S_{j , L (R)}^z$ and the $\xi_{d_b}$'s being 
pertinent scaling functions. Based on rather general assumptions, one expects some analog to 
Eq.~(\ref{icf.3}) to describe ${\bf \Sigma} [ x ]$ for $x / \xi_K > 1$, as well. Later, we 
provide a semiqualitative argument to infer how  ${\bf \Sigma} [ x ]$ behaves outside of 
the Kondo cloud. To exactly reconstruct  the scaling behavior encoded  in Eq.~(\ref{icf.3}), we employed DMRG approach to numerically 
evaluate ${\bf \Sigma} [ x ]$ in the case of a central impurity ${\bf S}_{\bf G}$, with either symmetric or 
non-symmetric couplings, as well as with a zero, or a nonzero, $B$ applied to ${\bf S}_{\bf G}$ (for the sake 
of presentation clarity, in the remainder of this section, we only discuss the symmetric, $B=0$ case. Later on 
in the paper, we consider the more general, non-symmetric situation).

To recover the scaling behavior of 
${\bf \Sigma} [ x ]$ and to eventually estimate $\xi_K$, we follow the strategy of Ref.~[\onlinecite{sholl}],
by making a combined use of the technique based on the scaling collapse of ${\bf \Sigma} [ x ]$ 
and of the technique based on the 
collapse of the Kondo length. To compare and combine the two strategies, in the 
following we devote two separate subsections to discuss the results obtained with the two techniques. As we
show below, to estimate $\xi_K$ it is enough to analyze scaling inside the Kondo cloud. For this reason, we 
mostly concentrate on the region characterized by $x / \xi_K \ll 1$ and briefly discuss at the end of the 
section the behavior of ${\bf \Sigma} [ x ]$ outside of the Kondo cloud. Eventually, we 
compare the final results with the ones obtained within the perturbative RG approach of 
appendix~\ref{reng}.

\subsection{The scaling collapse technique}
\label{scacol}

\begin{figure}[ht]
\centering
\includegraphics*[width=1.\linewidth]{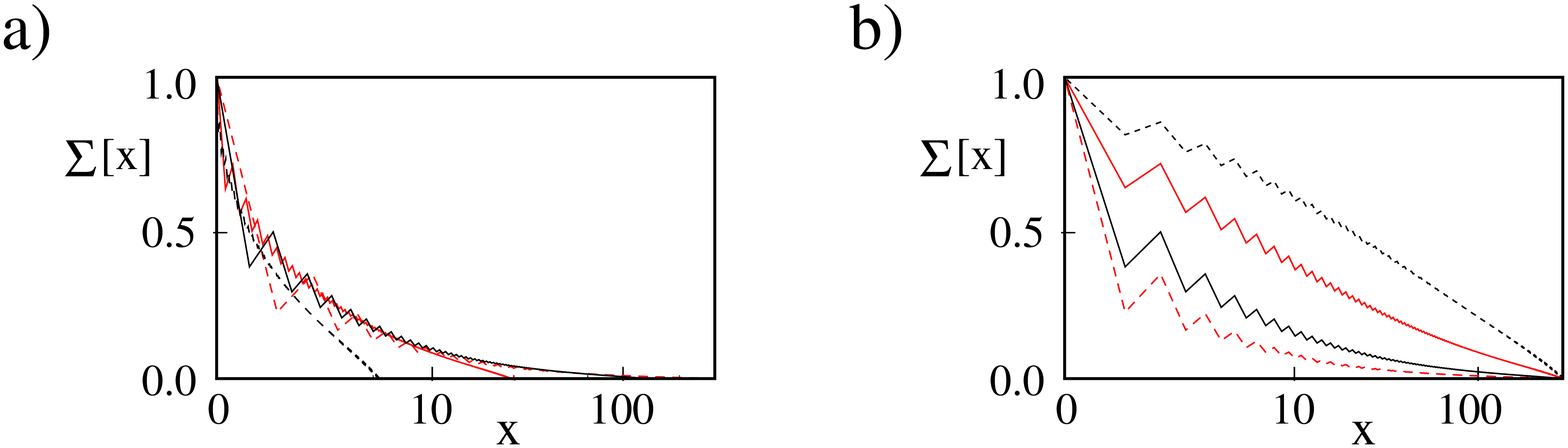}
\caption{\\
{\bf a):} Semilogarithmic rescaled curves for ${\bf \Sigma} [ x ] $    corresponding to $\ell = 300$, 
$\Delta = 0.3$ and, respectively,  $J' /J = 0.1$ (dashed black curve), $J'/J = 0.2$ (full red curve), $J' /J = 0.4$ (full black curve), 
and  $J'/J = 0.6$ (dashed red curve); \\
{\bf b):} Same as in panel {\bf a)}, but without rescaling. }
\label{0.3_scal}
\end{figure}

\begin{figure}[ht]
\centering
\includegraphics*[width=1.\linewidth]{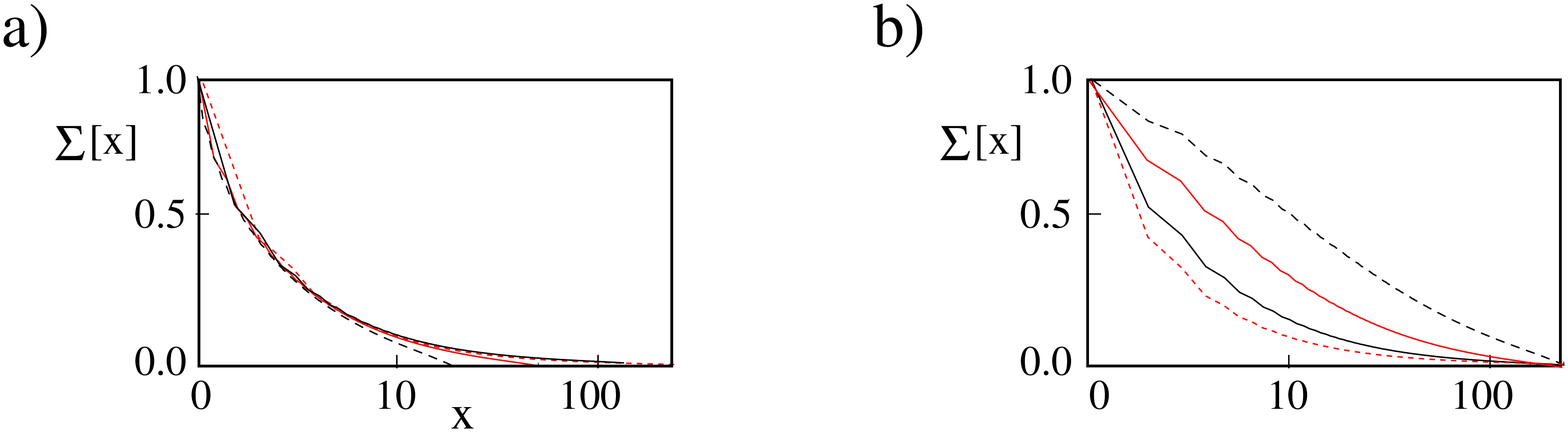}
\caption{\\
{\bf a):} Semilogarithmic rescaled curves for ${\bf \Sigma} [ x ] $    corresponding to $\ell = 300$, 
$\Delta = -0.3$ and, respectively,  $J' /J = 0.1$ (dashed black curve), $J'/J = 0.2$ (full red curve), $J' /J = 0.4$ (full black curve), 
and  $J'/J = 0.6$ (dashed red curve); \\
{\bf b):} Same as in panel {\bf a)}, but without rescaling. }
\label{m0.3_scal}
\end{figure}

The scaling collapse technique (SCT) is based on the expected scaling properties of ${\bf \Sigma } [ x ]$
in the limit in which $ x , \xi_K \ll \ell$. In this regime, Eq.~(\ref{icf.3}) reduces to 
\beq
{\bf \Sigma} [ x ] \approx  \sum_{ d_b } \ell^{d_b } \xi_{d_b} \left[ 0 ; \frac{x}{\xi_K} \right] 
\, . 
\label{icf.4}
\eneq
\noindent 
At a given $\ell$, Eq.~(\ref{icf.4}) shows that ${\bf \Sigma} [ x ]$ becomes a scaling function of $x / \xi_K$. 
Based on this observation, one readily concludes that, provided $\ell$ is large enough, 
curves for ${\bf \Sigma} [ x ]$ drawn 
at different values of the $J'$ (which means at different values of $\xi_K$), are expected, 
for $ x / \xi_K < 1$, to collapse onto
each other, provided $x$  is rescaled with the corresponding $\xi_K$. This is 
the hearth of SCT. In principle, at fixed $\Delta$, given two different values 
of $J'$, say $J'_1$ and $J'_2$, one may regard the scaling factor that makes the corresponding curves for 
${\bf \Sigma} [ x ]$ collapse onto each other, as a fitting parameter. Once it is 
properly estimated via a fitting procedure, it becomes equal to 
$\xi_K [ J'_1 / J , \Delta] / \xi_K [ J'_2 /J , \Delta ]$ (note that we 
henceforth denote with $\xi_K [ J' / J , \Delta ]$ the Kondo screening length 
at given $J_L^{'} = J_R^{'} = J' $ and $\Delta$). 
The RG approach of appendix~\ref{reng} provides us with a direct mean to analytically derive 
$\xi_K [ J' / J , \Delta ]$ up to an over-all factor independent of $J'$ and $\Delta$ determined by cutoff 
$\ell_0$. Yet, since any rescaling factor is given by the ratio between two screening lengths at different 
values of $J' / J$, it is always independent of the over-all factor. 
This enables us to directly  compare the DMRG results for the 
scaling factors obtained within SCT with the analytical results provided by RG approach. 
As we show below, the collapse of the Kondo length eventually lets one fix the over-all factor in $\xi_K$.  
Specifically, to check both cases of antiferromagnetic and ferromagnetic correlations in the leads, 
we apply SCT to a system with $\Delta = 0.3$, corresponding 
to $g \approx 0.83754$ (see appendix~\ref{sll} for details), and with 
$\Delta = - 0.3$ corresponding to $g \approx 1.24065$. In both cases we derive plots of ${\bf \Sigma} [ x ]$ 
at fixed  $J' / J = 0.1,0.2,0.4,0.6$ and at various values of $\ell$, the largest of which corresponds to 
$\ell = 300$. In Fig.~\ref{0.3_scal}{\bf a)} we plot, on a semilogarithmic scale on the $x$-axis, 
${\bf \Sigma} [ x ]$ vs. $x$ for $\ell = 300$, $\Delta = 0.3$, and for 
the values of $J' / J$ listed above. To compare DMRG results with the ones obtained within perturbative RG method, 
in drawing the plots we rescale $x$ by the scaling factors for the corresponding values of $J' / J$, determined 
using the formulas of appendix~\ref{reng} for $\xi_K$ and summarized in table~\ref{scalfact}. 
For comparison, in Fig.~\ref{0.3_scal}{\bf b)} we draw
the same plots, but without rescaling $x$. In Fig.~\ref{m0.3_scal}{\bf a)} and~\ref{m0.3_scal}{\bf b)}, 
we draw plots constructed following similar criteria, 
but now for  $\Delta = -0.3$. From the two figures, one clearly sees that, except for 
the black dashed curve in Fig.~\ref{0.3_scal}{\bf a)} (corresponding to 
the largest value of $\xi_K$ at $J'/J=0.1$ - we discuss this point in the following), 
the collapse is quite good.

\noindent

\noindent

\vspace{0.5cm}
 
\begin{table}
\centering
\begin{tabular}{| c | c |c|}
\hline 
${\rm Scaling \:  factor}$&$\Delta=0.3$&$\Delta=-0.3$\\
 \hline
$\frac{\xi_K [ 0.6 , \Delta]}{\xi_K [ 0.4 , \Delta ]}$ &  0.4554 & 0.5662 \\
 \hline
$\frac{\xi_K [ 0.6 , \Delta]}{\xi_K [ 0.2 , \Delta ]}$ &  0.0974 & 0.2006 \\
 \hline
$\frac{\xi_K [ 0.6 , \Delta]}{\xi_K [ 0.1 , \Delta ]}$ & 0.0181 &  0.0676 \\
 \hline 
\end{tabular}
\caption{Scaling factors for $\Delta= 0.3$ and $\Delta=-0.3$ and for $J' /J=0.6,0.4,0.2,0.1$ evaluated
using the scaling collapse technique.} 
   \label{scalfact}
\end{table}
 
An important observation about our method is that, differently from what has been 
done in Ref.~[\onlinecite{sholl}], we do not fit the ratios between the Kondo screening lengths from 
the numerical data. Instead, we compute them within perturbative  RG approach 
and eventually find that the collapse of the curves is quite good after rescaling $x$ with the values we 
computed. In Fig.~\ref{0.3_scal}{\bf a)}, we see quite 
a good collapse of the curves onto each other for any value of $J' / J$, but $J' / J = 0.1$. 
The lack of collapse in this last case can be traced back to a possible value of 
$\xi_K [ 0.1  , 0.3 ]$ exceeding the half-length of the chain ($\sim 300$). As we will show below, 
where we will be using a different technique allowing for directly estimating $\xi_K$, this is, in 
fact, the case, that shows the full consistency of our results with the expected Kondo scaling behavior.
At variance, in 
Fig.~\ref{0.3_scal}{\bf b)} we see a pretty good collapse of all the curves, implying that, in this case, 
all the Kondo screening lengths are $< 300$, including $\xi_K [ 0.1 , - 0.3 ]$. In 
addition, due to the fact that the bulk spin correlations are now ferromagnetic ($\Delta < 0$), the 
staggered component of the integrated spin correlations disappears, all the curves look quite smooth,
as a function of $x$, and the corresponding collapse is even more evident than that of Fig.~\ref{0.3_scal}.

To ultimately fix the over-all factor in $\xi_K$, we now discuss the Kondo length collapse 
technique.

\subsection{The Kondo length collapse technique}
\label{klc}

\begin{figure}[ht]
\centering
\includegraphics*[width=.98\linewidth]{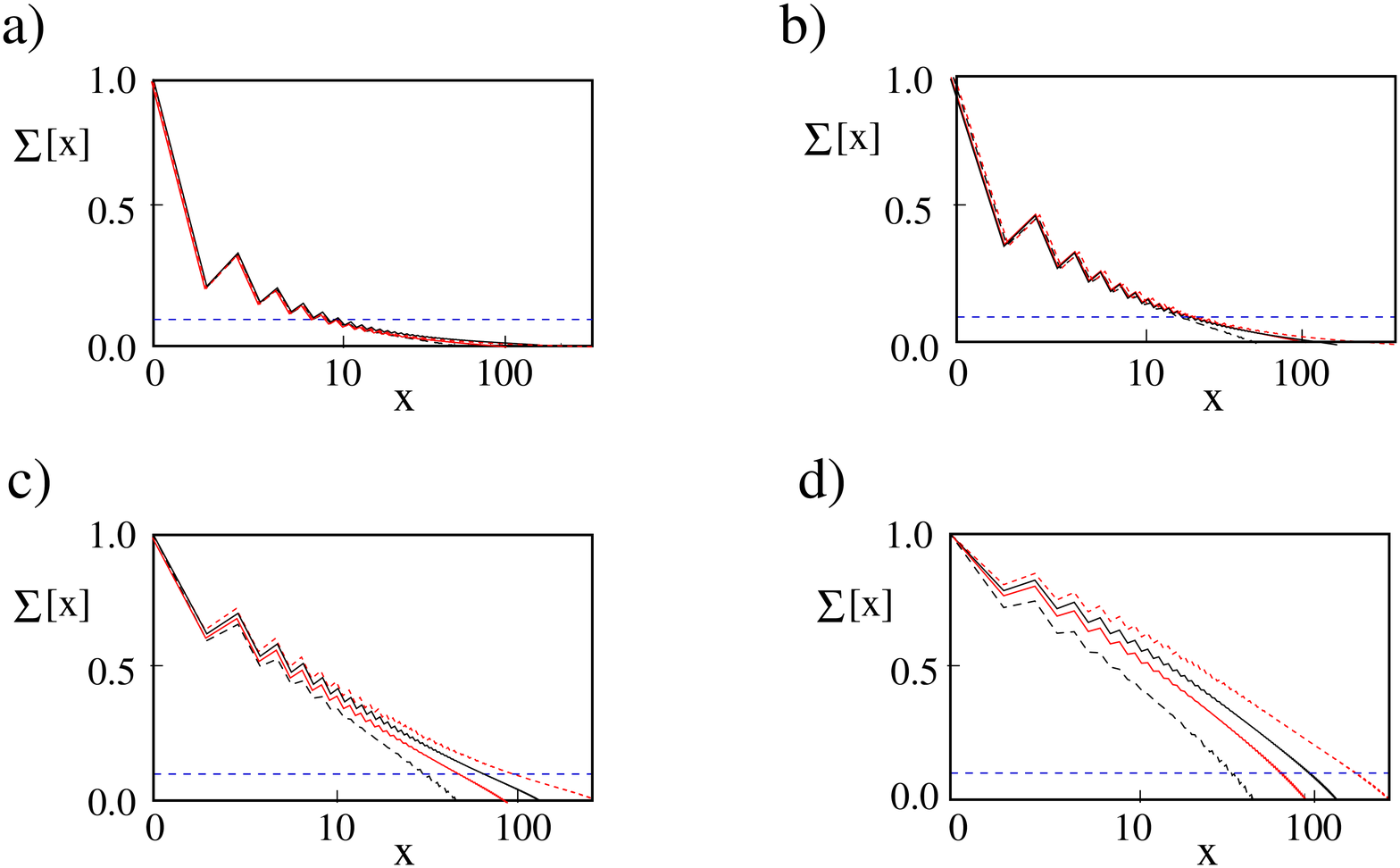}
\caption{\\
{\bf a):} Curves for ${\bf \Sigma} [ x ] $ at $\Delta = 0.3$,  $J'/J = 0.6$ and   $\ell = 50$  (dashed black curve),
$\ell = 100$ (solid red curve), $\ell = 150$ (solid black curve), $\ell = 300$  (dashed red curve). As a guide to the eye, the horizontal line at $y=0.1$ is shown as a dashed blue segment; \\
{\bf b):} Same as in panel {\bf a)}, but for $J'/J=0.4$; \\
{\bf c):} Same as in panel {\bf a)}, but for $J'/J=0.2$; \\
{\bf d):} Same as in panel {\bf a)}, but for $J'/J=0.1$.  }
\label{klsca_0.3}
\end{figure}

The Kondo length collapse technique (KLCT) is grounded on the ``physical'' meaning of the 
Kondo cloud as the cloud of spins fully screening ${\bf S}_{\bf G}$
into the Kondo singlet  \cite{affleck_length}.  In the presence of a perfect screening,  one would  expect 
$\xi_K$ to emerge as the first zero of ${\bf \Sigma} [ x ]$ one meets when moving from 
the impurity location into the leads. In practice, as we discuss above, the actual zero of 
${\bf \Sigma} [ x ]$ is set at $x = \ell$ by the over-all boundary conditions. Therefore, to 
extract $\xi_K$ one first of all sets a conventional ``reduction factor'' $r (<1)$ by 
defining a putative Kondo screening length $\xi_K^{(r)}$ as the value of $x$ 
at which ${\bf \Sigma} [ x ]$ is reduced by $r$ with respect to 
its value at $x=0$, that is, ${\bf \Sigma} [ x = \xi_k^{(r)} ] = r$. 
Choosing a specific value for $r$ is equivalent to fix $\ell_0$. 
Yet, variations around a reasonable choice of $r$ (such that basically all, 
or almost all, of the Kondo cloud resides over distances $\leq \xi_K^{(r)}$ from the 
impurity location) just affect the estimated value of $\xi_K^{(r)}$ by a factor of order 1 \cite{sholl}. 
Thus, in the following we choose to follow Ref.~[\onlinecite{sholl}], by choosing 
$r = 0.1$ and accordingly using $\xi_K^{(0.1)}$ evaluated at given $\Delta$ and $J' / J$ as 
an estimate of $\xi_K [ J' / J , \Delta ]$. This eventually allows us to uniquely set 
$\ell_0$ and to provide the actual values of $\xi_K$  for various 
choices of the system parameters. In practice,  at a given $\ell$, 
fixing $\Delta$ and $J' / J$, one uses DMRG results to extract 
an $\ell$-dependent scale $\xi_K^{(0.1)}  [ J' / J , \Delta , \ell ]$ by means of the 
condition ${\bf \Sigma} [ x = \xi_K^{(0.1)}  [ J' / J , \Delta , \ell ] ] = 0.1$. For large 
enough values of $\ell$,  $\xi_K^{(0.1)}  [ J' / J , \Delta , \ell ]$ is expected to 
reach an asymptotic value $\xi_K^{(0.1)}  [ J' / J , \Delta ]$ which is independent of 
$\ell$ and, according to the above observations, provides the corresponding estimate of $\xi_K$. 
Based on these observations, in Fig.~\ref{klsca_0.3} we plot ${\bf \Sigma} [ x ] $ vs. $x$ on a 
semilogarithmic scale (on the $x$ axis), at $\Delta = 0.3$ and,  respectively, 
$J'/J = 0.6$ (Fig.~\ref{klsca_0.3}{\bf a)} ), $J'/J = 0.4$ (Fig.~\ref{klsca_0.3}{\bf b)} ), 
$J'/J = 0.2$ (Fig.~\ref{klsca_0.3}{\bf c)} ), $J'/J = 0.1$ (Fig.~\ref{klsca_0.3}{\bf d)} ). 
All the plots display curves corresponding to $\ell = 50$ (dashed black curve), $\ell=100$ (solid red curve), 
$\ell=150$ (solid black curve),and $\ell=300$ (dashed red curve).

According to the discussion above, from Fig.~\ref{klsca_0.3}{\bf a)} and from Fig.~\ref{klsca_0.3}{\bf b)} 
we conclude that both $\xi_K [ 0.6 , 0.3]$ and $\xi_K [ 0.4 , 0.3]$ are $\ll 150$. In fact, a numerical 
estimate provides $\xi_K [ 0.6, 0.3] \approx 10.23$ and  $\xi_K [ 0.4 , 0.3] \approx 23.11$. At variance, 
the absence of collapse at ${\bf \Sigma} [ x ] = r$ for $J' / J = 0.1$ and $=0.2$ implies that 
in both cases $\xi_K$ must be comparable with (or larger than) $\ell = 150$. 
Knowing the actual value of $\xi_K [ 0.6, 0.3]$ allows us to estimate $\xi_K [ 0.2 , 0.3]$ and $\xi_K [ 0.1 , 0.3]$
by just using the scaling ratios derived in section~\ref{scacol} within the SCT. 
The results are summarized in table~\ref{kles}.  Apparently, they confirm 
the conclusion that both  $\xi_K [ 0.2 , 0.3]$ and $\xi_K [ 0.1 , 0.3]$ are 
$ > 150$. The shorter values 
of the Kondo screening length at a given $J' / J$ (compared to the ones at $\Delta = 0.3$) allow us 
to directly estimate $\xi_K [ J' / J , - 0.3 ]$ for $J' / J = 0.6 , 0.4 , 0.2$. This time, only 
$\xi_K [ 0.1 , - 0.3 ]$ had to be found using the corresponding scaling ratio derived in section~\ref{scacol}.

\noindent

\begin{figure}[ht]
\centering
\includegraphics*[width=.98\linewidth]{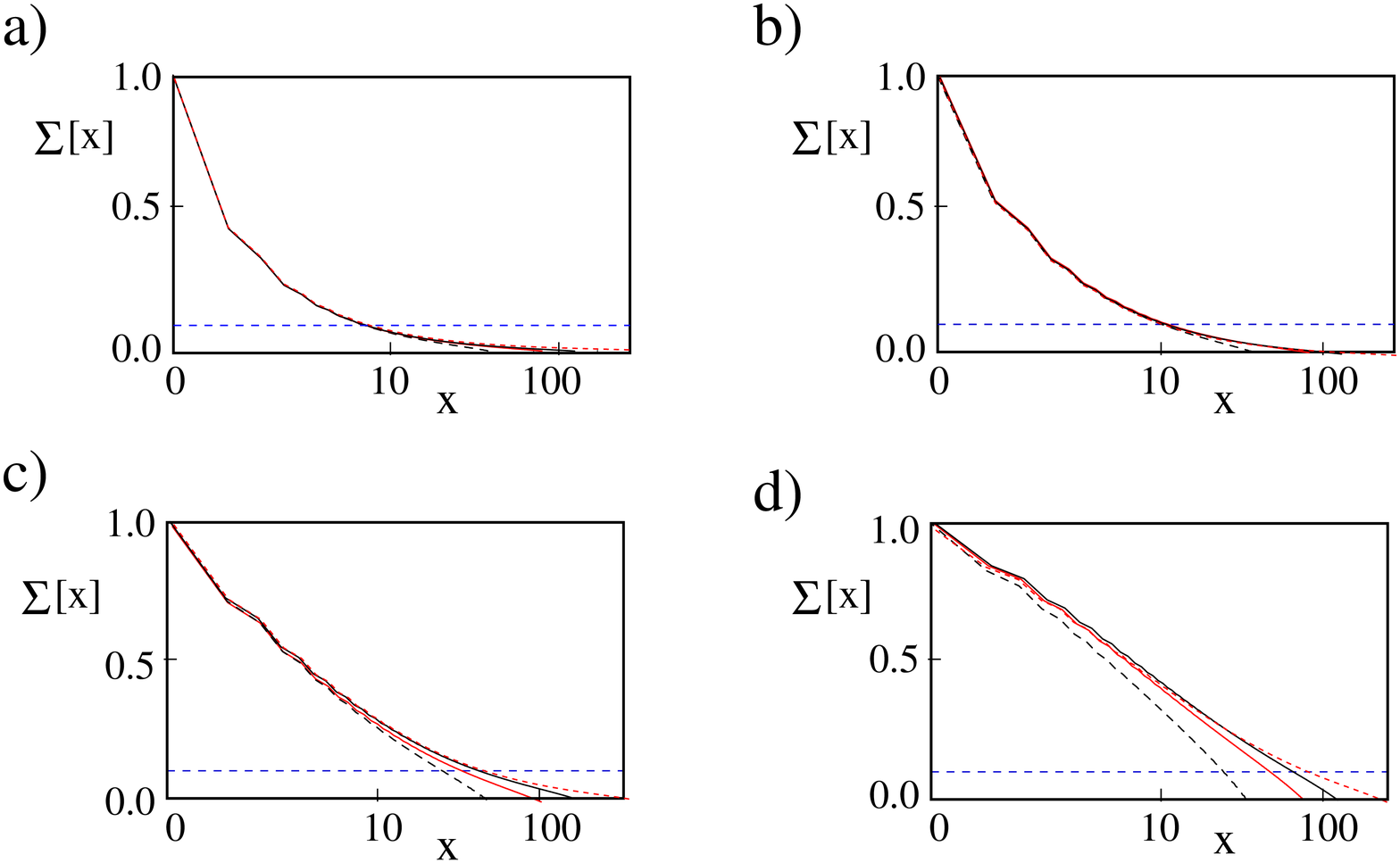}
\caption{\\
{\bf a):} Curves for ${\bf \Sigma} [ x ] $ at $\Delta = -0.3$,  $J'/J = 0.6$ and   $\ell = 50$  (dashed black curve),
$\ell = 100$ (solid red curve), $\ell = 150$ (solid black curve), $\ell = 300$  (dashed red curve). As a 
guide to the eye, the horizontal line at $y=0.1$ is shown as a dashed blue segment; \\
{\bf b):} Same as in panel {\bf a)}, but for $J'/J=0.4$; \\
{\bf c):} Same as in panel {\bf a)}, but for $J'/J=0.2$; \\
{\bf d):} Same as in panel {\bf a)}, but for $J'/J=0.1$.  }
\label{klsca_m0.3}
\end{figure} 
\noindent

To stress the possibility of estimating some Kondo lengths by only combining the Kondo 
length collapse with the scaling collapse approach, in table~\ref{kles} we report in 
black the values directly estimated using KLCT, in red the ones inferred combining 
KLCT with the results of section~\ref{scacol} for the scaling factors. 
 
\begin{table}
\centering
\begin{tabular}{| c | c |c|}
\hline 
$J' / J$ & $\xi_K [ J' / J , 0.3 ] $&$ \xi_K [ J' / J,  -0.3 ]$\\
\hline
 0.6  &  10.23 & 9.61 \\
\hline
0.4 &  23.11 & 16.47 \\
\hline
0.2 & {\color{red} 109.14} &  48.33 \\
\hline 
0.1 & {\color{red} 565.19} & {\color{red} 142.16} \\
\hline
\end{tabular}
\noindent
\caption{Values for the Kondo length for $\Delta= 0.3$ and $\Delta = -0.3$ and for $J'/J=0.6,0.4,0.2,0.1$ 
evaluated using the KLCT (numbers displayed in black), or combining the KLCT  with the SCT (numbers displayed in red).}
 \label{kles}
\end{table}

\vspace{0.5cm}
As a general, concluding comment about KLCT, we note 
that, in order for the method to be effective, we need at least the two curves 
corresponding to the largest value of $\ell$ and to the next-to-largest one ($\ell^{'}$)
to collapse onto each other. Since this implies that both of 
them must not be affected by finite-size effect, 
we infer that the necessary condition for the collapse to happen is that $\xi_K \ll \ell^{'} $,
which motivates the absence of collapse in some of the plots in Fig.~\ref{klsca_0.3}
and in Fig.~\ref{klsca_m0.3}.  Of course, 
one could increase $\ell$ and directly estimate the value of $\xi_K$ from the collapse. Yet, 
for the sake of the presentation, we prefer to present some plots not showing collapse, in 
order to be able to discuss the main scenario and to show the remarkable consistency of the 
exact numerical data with the analytical results obtained within SLL framework, even 
for chains with a limited number of sites. 
Note that, after fitting the value of $\ell_0$ from DMRG results, the perturbative RG equations
provide quite good estimates for $\xi_K [ J' / J , \Delta ]$ and can be effectively used for 
such a purpose as, for instance, is was done in Ref.~[\onlinecite{gst}]. 

To conclude the discussion of the fully symmetric system, we now briefly comment on 
the behavior of ${\bf \Sigma} [ x ]$ outside of the Kondo cloud.  
  
\subsection{Kondo screening cloud in the XXZ spin chain}
\label{kscalut}

The way we apply SCT and KLCT to obtain $\xi_K$ from DMRG data relies upon the 
validity of Eq.~(\ref{icf.3}) inside the Kondo cloud. Yet, based on very general grounds, 
a scaling form for ${\bf \Sigma} [ x ]$ such as the one in Eq.~(\ref{icf.3}) is 
expected to apply outside of the Kondo cloud, as well, provided $x , \xi_K \ll \ell$ 
\cite{affleck_length,barzykin,baraf} though, clearly, the perturbative RG estimate of the 
right-hand side of Eq.~(\ref{icf.3}) and of  Eq.~(\ref{wcf.0a}), does no more apply. To pertinently 
replace Eq.~(\ref{wcf.0a}) we need the analog, for our spin-chain model, of the conformal field
theory based nonperturbative approach to Kondo screening cloud developed by 
Affleck and Ludwig \cite{affleck_1,ludwig_1}. 
To do so, we have to work out the analog, in our case,  
of Nozier\`es Fermi liquid theory \cite{noz_1,noz_2}. In fact, 
in the spin chain framework, the analog of Nozier\`es Fermi liquid is the ``healing'' of the chain, that is, 
the saturation of the running couplings to values that are of the order of all the other bulk couplings \cite{gst}. 
In addition, the Kondo cloud emerges around ${\bf S}_{\bf G}$ of size $\sim 2 \xi_K$. 
This can be roughly regarded as an extended region 
${\cal R}$ embedded within the chain, which is coupled at its endpoints to the spins in the remaining part 
of the chain by means of the boundary Hamiltonian $H_{\rm SC}$, given by 
\beq
H_{\rm SC} = J \{ S_{\xi_K , L}^+ S_{\xi_K+1 , L}^- + S_{\xi_K , L }^- S_{\xi_K+1 , L }^+ + \Delta S_{\xi_K , L }^z S_{\xi_K+1 , L}^z \}
+  J \{ S_{ \xi_K , R}^+ S_{ \xi_K + 1 , R}^- + S_{ \xi_K , R}^- S_{ \xi_K + 1 , R}^+ + \Delta S_{\xi_K , R}^z S_{ \xi_K + 1 , R}^z \}
\, . 
\label{ww.1}
\eneq
\noindent
Regarding the Kondo cloud as  a spin-singlet spin cluster of size $\sim 2 \xi_K$ 
coupled to two spin chains at its endpoints allows to employ a pertinent generalization 
of the derivation in appendix~\ref{wlink} to recover the behavior  of 
${\bf \Sigma} [ x ]$ for $x > \xi_K$.  First, we note that, due to the boundary condition 
${\bf \Sigma} [ \ell ] = 0$, we may equivalently set 
\beq
{\bf \Sigma}  [ x ] =   - \sum_{i = x+1}^{\ell} \sum_{X = L , R}  \left\{ \frac{\langle S_{\bf G}^z S_{i , X}^z \rangle - 
\langle S_{\bf G}^z \rangle \langle S_{i , X}^z \rangle }{
\langle (S_{\bf G}^z )^2 \rangle - ( \langle S_{\bf G}^z \rangle )^2 } \right\}  
\, . 
\label{icf.xx1}
\eneq
\noindent
Therefore, we see that, due to strong singlet correlations within the Kondo 
cloud, one may legitimately approximate the whole chain ground state as 
$ | \Psi \rangle_0 = | {\rm KC} \rangle \otimes 
| {\bf 0} \rangle_L \otimes | {\bf 0} \rangle_R$, with $ | {\rm KC} \rangle$ being the 
``Kondo cloud spin singlet ground state'' and $| {\bf 0} \rangle_L $ and $ | {\bf 0} \rangle_R$ respectively being the ground states of 
the portion of the $_L$ and of the $_R$ spin chain ranging from $\xi_K + 1 $ to $ \ell $. Therefore,   
due to the singlet nature of $ | {\rm KC} \rangle$, one obtains $\langle \Psi_0 | S_{\bf G}^z S_{j , L (R)}^z
 | \Psi_0 \rangle = 0$ whenever $j > \xi_K$. Accordingly, 
to estimate the leading nonzero contribution to the correlation function, one has to 
correct the system's ground state from $ | \Psi_0 \rangle$ to a state $ | \Psi_1 \rangle $ taking into account the 
effects of $H_{\rm SC}$ in Eq.~(\ref{ww.1}). To leading order, one obtains 
\beq
| \Psi_1  \rangle\approx | \Psi \rangle_0 - \left\{ J \Delta S_{\xi_K + 1 , L}^z \sum_X \: | X  \rangle \frac{ \langle X | S_{\xi_K , L}  | \Psi \rangle_0 }{  
\delta E_X } 
+ J \Delta S_{\xi_K + 1 , R}^z \sum_X \: | X  \rangle \frac{ \langle X | S_{\xi_K , R}  | \Psi \rangle_0 }{  \delta E_X } \right\}
 \otimes | {\bf 0} \rangle_L \otimes | {\bf 0} \rangle_R \, , 
\label{ww.2}
\eneq
\noindent
with the sum at the right-hand side of Eq.~(\ref{ww.2}) taken over low-lying excited states of 
the Kondo cloud spin 
singlet, $\{  | X \rangle \}$, with corresponding excitation energy $\delta E_X $ (measured with 
respect to the ground state). As a result, to leading order in $H'$, one obtains for $j > \xi_K$
\begin{eqnarray}
&& \frac{ \langle \Psi_1 | S_{\bf G}^z S_{j , L ( R ) }^z | \Psi_1 \rangle  - \langle \Psi_1 | S_{\bf G}^z | \Psi_1 \rangle 
\langle \Psi_1 | S_{j , L ( R ) }^z | \Psi_1 \rangle }{ \langle \Psi_1 | ( S_{\bf G}^z )^2  |  \Psi_1 \rangle
- ( \langle \Psi_1 |  S_{\bf G}^z  |   \Psi_1 \rangle )^2 } \approx \nonumber \\ 
&& - J \Delta  \sum_X \left\{ \frac{ \langle \Psi_0 | S_{\xi_K , L ( R)}^z | X \rangle \langle X | S_{\bf G}^z | \Psi_0 \rangle }{\delta E_X } 
\right\} \: _{L ( R )  } \langle {\bf 0} | S_{\xi_K + 1 , L (R)}^z S_{j , L (R)}^z | {\bf 0} \rangle_{L ( R ) } 
\,. 
\label{ww.3}
\end{eqnarray}
\noindent
To compute the correlation functions  
$_{L ( R )  }  \langle {\bf 0} | S_{\xi_K + 1 , L (R)}^z S_{j , L (R)}^z | {\bf 0} \rangle_{L ( R ) } $, we 
use the result for the homogeneous open XXZ chain, Eq.~(\ref{corr.x2}), by substituting 
$\ell$ with $\hat{\ell} = \ell - \xi_K$ and $x$ with $\hat{j} = j - \xi_k$. As a result, we therefore
conclude that these correlation functions are independent of $\xi_K$ up to terms $\propto ( \xi_K / \ell ) \ll 1$.
Moreover, at a given $ | X \rangle$, the 
matrix element $\langle X | S_j^z | \Psi_0 \rangle$ is poorly dependent on $j$, as long as the spin ${\bf S}_j$ lies 
within the Kondo cloud. Since one expects $\delta E_X \sim \xi_K^{-1}$, we eventually 
combine Eqs.~(\ref{icf.xx1}, \ref{ww.2}, \ref{ww.3}) to conclude that, for $x > \xi_K$, one obtains 
\beq
{\bf \Sigma} [ x ] \approx \xi_K \sum_{\bar{d}_b} \: \ell^{\bar{d}_b} \: 
 \omega_{\bar{d}_b}  \left( \frac{x}{\ell} \right) + \ldots 
 \, , 
 \label{ww.4}
 \eneq
\noindent
with the sum taken over a pertinent set of scaling exponents and the ellipses standing for additional contributions 
$\propto \xi_K / \ell$, which we neglect, due to the assumed condition $\xi_K / \ell \ll 1$. For instance, from 
Eq.~(\ref{corr.x2}), we infer that, for $\xi_K \ll x \ll \ell$, one obtains 
\beq
   \sum_{\bar{d}_b} \: \ell^{\bar{d}_b} \: 
 \omega_{\bar{d}_b}  \left( \frac{x}{\ell} \right)  \approx 
 - \frac{g}{2 \ell^2} \: \sum_{j = x+1}^\ell \left[ \frac{1}{1 - \cos \left( \frac{\pi j}{\ell} \right)} \right] 
  - \frac{2 a g}{\ell} \:  \sum_{j = x+1}^\ell \left[ (-1)^j \: \left| \frac{2 \ell}{\pi} \sin \left( \frac{\pi j}{\ell}
 \right) \right|^{-g} \cot \left( \frac{\pi j}{2 \ell} \right) \right] 
 \, . 
 \label{ww.5}
 \eneq
\noindent 
As a result, we expect that, when synoptically considering plots of ${\bf \Sigma} [ x ]$ 
derived at different values of $J' / J$ (that is, of $\xi_K$), with $x$ lying outside of the Kondo cloud, 
the curves collapse onto each other, provided ${\bf \Sigma} [x ]$ is rescaled to 
$\hat{\bf \Sigma} [ x ] = \xi_K^{-1} {\bf \Sigma} [ x ]$, which incidentally appears to 
be consistent with Affleck-Ludwig result for 
the real-space correlations at $x$ lying outside of the Kondo cloud \cite{affleck_1,ludwig_1}.

\begin{figure}[ht]
\centering
\includegraphics*[width=1.\linewidth]{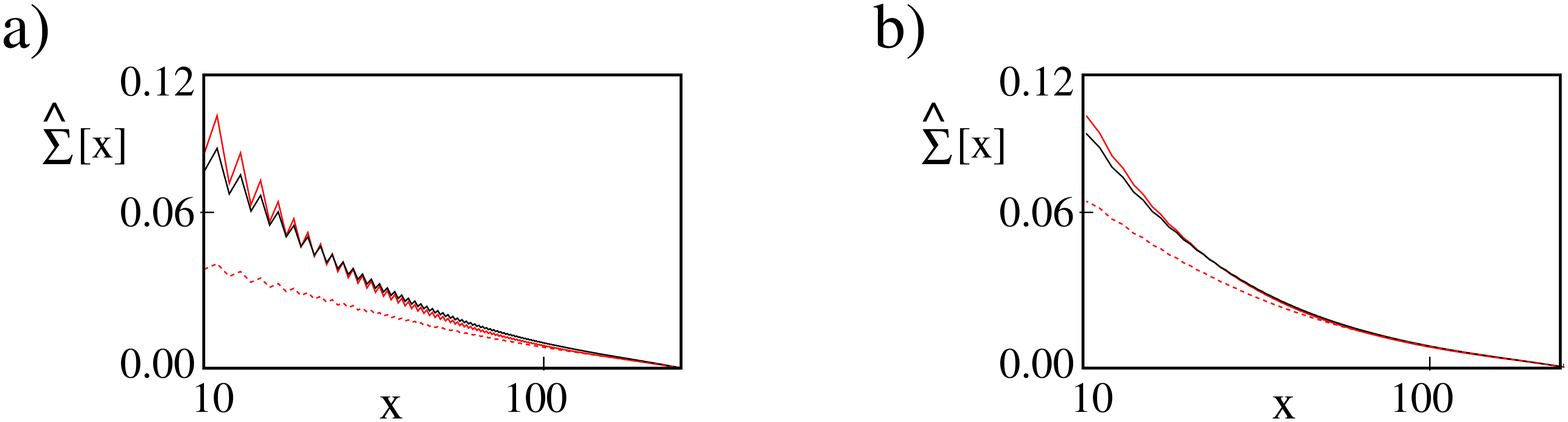}
\caption{\\
{\bf a):} Rescaled curves (see main text)  for $\hat{\bf \Sigma} [ x ] $ at $\Delta =
 0.3$,  $J'/J = 0.6$ and   $\ell = 300$  (dashed red curve),
$J' / J = 0.4$ (solid black curve), and $J' / J = 0.2$ (solid red curve). 
There is an apparent collapse of the first two curves onto each other for 
$x > \xi_K [ 0.4 , 0.3]$ and of all three  curves onto each other for $x > \xi_K [ 0.2 , 0.3 ]$;  
{\bf b):} Same as in panel {\bf a)}, but for $\Delta = -0.3$.  }
\label{scaly}
\end{figure}

To check this result, in Fig.~\ref{scaly} we plot curves for $\hat{\bf \Sigma} [ x ]$ for $\Delta = 0.3$ and 
$J' / J = 0.2,0.4,0.6$ (Fig.~\ref{scaly}{\bf a)}), and for $\Delta = -0.3$ and the same values of 
$J' / J$ (Fig.~\ref{scaly}{\bf b)}). In drawing the plots, we rescaled the various curves with the 
ratios between the corresponding Kondo screening lengths, as derived in Sec.\ref{scacol}. 
In the two plots, the dashed red curve and the solid black curve respectively correspond to 
$J' / J = 0.6$ and $J' / J = 0.4$. 
We see that, both for $\Delta = 0.3$ and for $\Delta = -0.3$, the two curves collapse onto each other for 
$x > \xi_K [ 0.4 , \Delta ]$. Moreover, in both plots we note that the solid red curve (corresponding to 
$J' / J =0.2$) collapses onto the other two ones for $x > \xi_K [ 0.2 , \Delta ]$. This remarkable result 
is consistent with the discussion provided above and constitutes another direct evidence 
for the emergence of the Kondo cloud over a length scale $\sim \xi_K$. 

\noindent
We now move to discuss models in which either the $_L$-$_R$-symmetry in the Kondo couplings, or 
the spin-parity symmetry (or both) are broken and see how the lack of those symmetry 
affects the main picture for the Kondo cloud we derived so far.

\section{Non symmetric Kondo interaction Hamiltonian}
\label{nonsymmetric}

In this section we discuss how Kondo effect is affected by either a breaking of
the symmetry between the couplings of ${\bf S}_{\bf G}$ to the two leads, or 
by the onset of a nonzero $B$ (or both), starting with   
the asymmetry in the couplings to the leads.
 
\subsection{Magnetic impurity with asymmetric Kondo couplings}
\label{asym}

\begin{figure}[ht]
\centering
\includegraphics*[width=1.\linewidth]{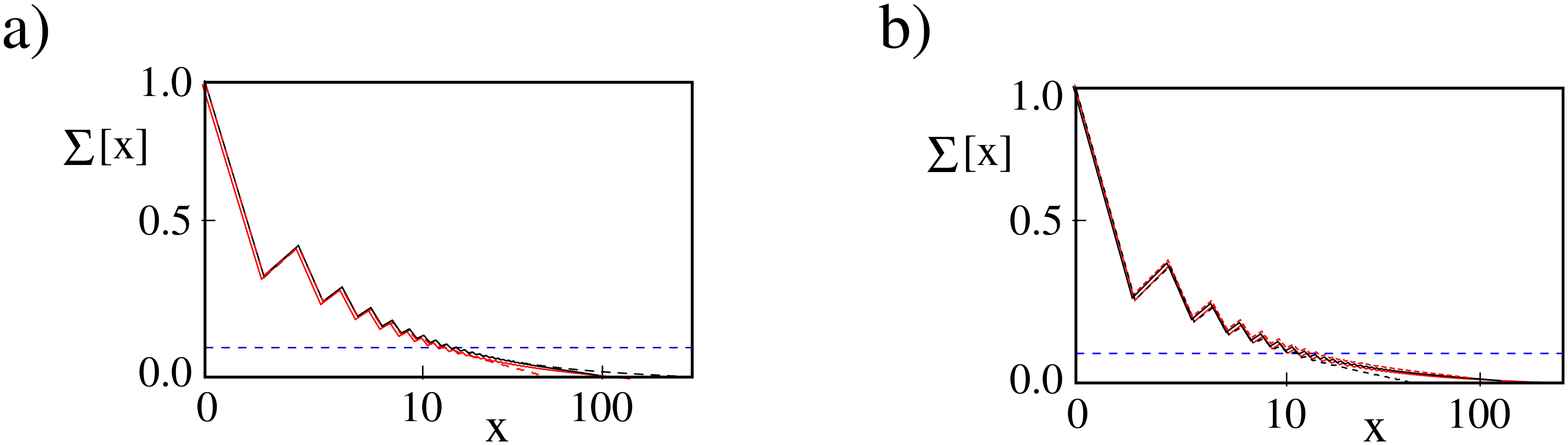}
\caption{\\
{\bf a):} Curves for ${\bf \Sigma} [ x ] $ at $\Delta = 0.3$,  $J_R^{'}/J = 0.6$, $J_L^{'} / J = 0.4$, and   $\ell = 50$  (dashed black curve),
$\ell = 100$ (solid red curve), $\ell = 150$ (solid black curve), $\ell = 300$  (dashed red curve). As a 
guide to the eye, the horizontal line at $y=0.1$ is shown as a dashed blue segment; \\
{\bf b):} Same as in panel {\bf a)}, but for $J_L^{'}/J=0.2$.  }
\label{asym_1}
\end{figure}

To discuss the effects of asymmetries in the Kondo couplings, here we assume that, in 
 ${\cal H}$ in  Eq.~(\ref{modham.1}), we have    $J_L^{'} < J_R^{'}$ and 
$J_{z , L ( R)}^{'} = \Delta J_{ L (R)}^{'}$,  which eventually implies 
$J_{z , L}^{'} < J_{z , R}^{'}$, as well.   
According to the discussion of section~\ref{klc} 
we again expect a Kondo cloud to emerge, with $\xi_K = \xi_{K , R}$, that is, 
$\xi_K$ is set by the stronger coupling of ${\bf S}_{\bf G}$, while 
the weaker coupling leads  to the  residual  Hamiltonian $H_W$ in Eq.~(\ref{mh.15}). As stated in section~\ref{klc},
$H_W$ does not lead to an additional dynamical length scale. In fact, it merely affects the value of 
$\xi_K$  by continuously renormalizing it from what one would have for 
just a magnetic impurity Kondo-coupled at the endpoint of a single XXZ chain \cite{eggert92}, to the value 
one obtains in the case of symmetric couplings.  Note that the dependence of $\xi_K$ on 
the (stronger) Kondo coupling strength can be readily inferred from  
Eqs.~(\ref{rengu.2}), which, just as in the symmetric case, fixes the screening length, up to 
an over-all factor.  The latter carries information on how the screening is distributed throughout the two channels
and, in general, it can hardly be recovered within SLL-based perturbative RG approach.
Thus, in the following we directly determine it from numerical DMRG data. Specifically,   to spell this point out, we 
define integrated correlation functions at both sides of ${\bf S}_{\bf G}$, ${\bf \Sigma}_{L , R } [ x ]$, 
both depending on a parameter $\chi$, so that 
\begin{eqnarray}
  {\bf \Sigma}_L [ x ] &=&  \chi +  \sum_{j = 1}^x \left\{ \frac{\langle S_{\bf G}^z S_{j , L}^z \rangle -\langle S_{\bf G}^z \rangle \langle S_{j , L}^z \rangle }{
 \langle (S_{\bf G}^z )^2 \rangle - ( \langle S_{\bf G}^z \rangle )^2 } \right\} \,, \nonumber \\
  {\bf \Sigma}_R [ x ] &=& 1 -\chi + \sum_{j = 1}^x \left\{ \frac{\langle S_{\bf G}^z S_{j , R}^z \rangle -\langle S_{\bf G}^z \rangle \langle S_{j , R}^z \rangle }{
 \langle (S_{\bf G}^z )^2 \rangle - ( \langle S_{\bf G}^z \rangle )^2 } \right\} 
 \, . 
 \label{n2c.6}
\end{eqnarray}
\noindent
By definition, from Eq.~(\ref{n2c.6}) one has ${\bf \Sigma} [ x ] = {\bf \Sigma}_L [ x ] + {\bf \Sigma}_R [ x ]$, 
which implies the boundary condition  ${\bf \Sigma} [ \ell ] = {\bf \Sigma}_L [ \ell ] +  {\bf \Sigma}_R [ \ell ] = 0$.
In addition, to fix $\chi$ we explicitly require that both integrated correlation functions are vanishing at 
$x = \ell$, that is, ${\bf \Sigma}_L [ \ell ] =   {\bf \Sigma}_R [ \ell ] = 0$. Doing so, 
we roughly state that
the Kondo cloud is, in general, non symmetrically distributed across the leads and is such that 
the part at the right(left)-hand side lead screens the impurity by a fraction equal to $1-\chi$ ($\chi$).
Apparently, one has $0 \leq \chi \leq 1$ and, varying $\chi$, one can continuously move from 
the two-channel Kondo regime we studied in section~\ref{icf}, 
corresponding to $\chi = 1/2$, in which the Kondo cloud is 
symmetrically distributed over the two leads, to the perfect one-channel Kondo regime, either 
corresponding to $\chi=0$, or to $\chi=1$, in which the Kondo cloud 
is fully distributed over one lead only.   $\chi$  appears to 
be a smooth function interpolating between the values 0 (at $J'_L / J_R^{'} = 0$) and 
$1$ (at $J'_L / J_R^{'} \to \infty$), and equal to $1/2$ at $J'_L / J_R^{'} = 1$. We also have $\chi < (>) 1/2$ 
according to whether $J_L^{'} < (>) J_R^{'}$, which implies that the lead that actually sets $\xi_K$ has to 
screen an impurity effectively larger by $1/2 - \chi$ ($\chi - 1/2$) than what it would be at the symmetric point. 
Eventually, this shows us the rationale of introducing Eqs.~(\ref{n2c.6}), that is, that 
any asymmetry between the Kondo couplings must imply an increase in $\xi_K$ with respect to the value it 
takes at the symmetric point. To check our prediction, in the following we estimate $\xi_K$ and $\xi_{K , R}$ 
for $\Delta = 0.3$, $J^{'}_R = 0.6$, $J_{z ,L (R)}^{'} = \Delta J_{L (R)}^{'}$, and 
$J^{'}_L /  J = 0.6,0.4,0.2$ (note that our choice for $J^{'}_R / J$ is expected, based on the results 
of the previous sections, to make $\xi_K$ of the order of 10 lattice spacings, which allows us to 
make reliable simulations using chain with at most 300 sites at each side of ${\bf S}_{\bf G}$). 
To estimate $\xi_K$, we used KLCT  at $J^{'}_L /  J = 0.4,0.2$. In Fig.~\ref{asym_1}, we show the collapse of the  
curves for ${\bf \Sigma} [ x ]$ derived  at  $\ell = 50,100,150,300$. The estimated values of $\xi_K$ are 
reported in table~\ref{estimated}. 

\begin{figure}[ht]
\centering
\includegraphics*[width=0.45\linewidth]{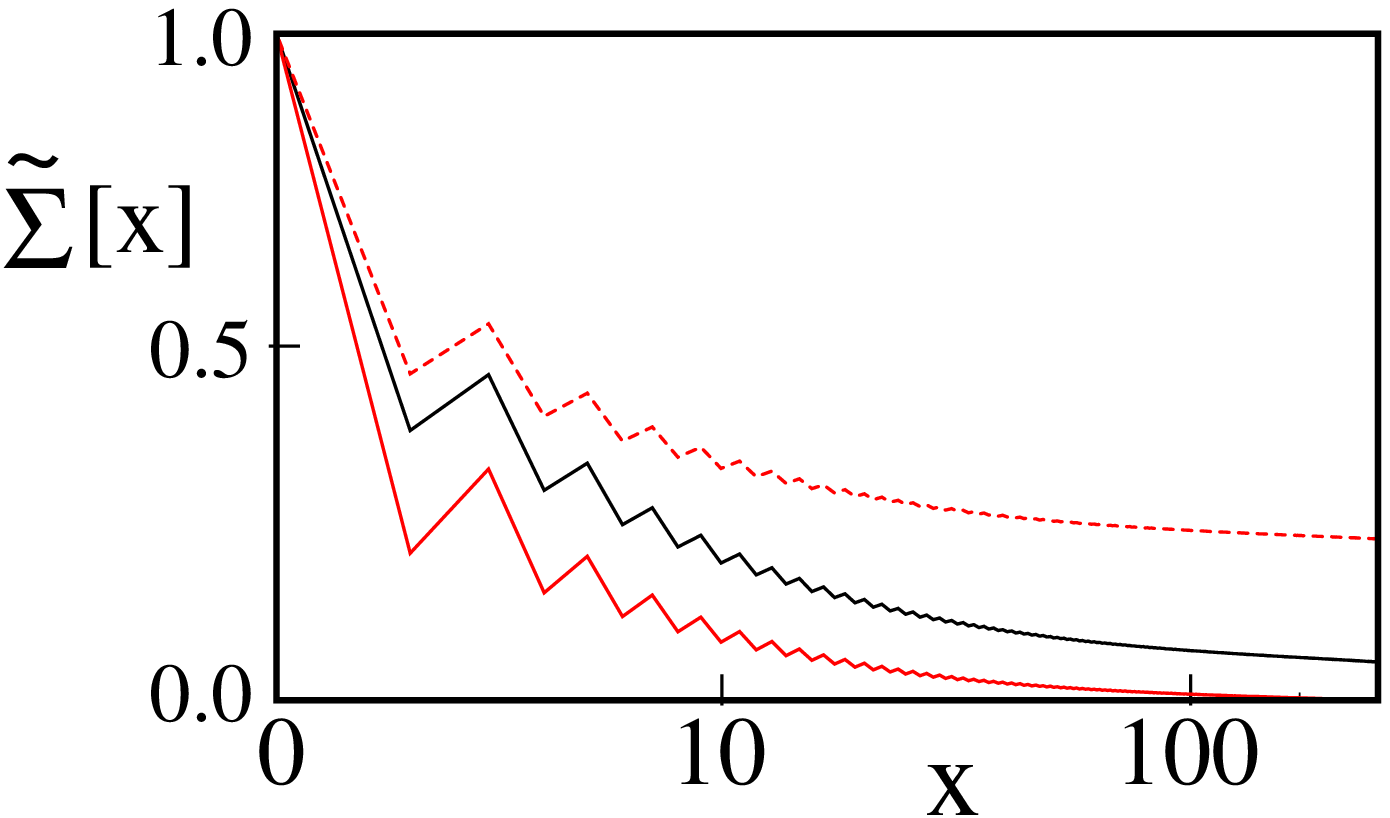}
\caption{Curves for $\tilde{\bf \Sigma}  [ x ] $ {\it vs.} $x$  at $\Delta = 0.3$,  $J_R^{'}/J = 0.6$, $\ell=300$, $J_L^{'} / J = 0.4$ (solid red curve), and    $J_L^{'} / J = 0.2$ 
(solid black curve). The estimated 
value of $\chi$ (see text) is $\chi =0.238$ in the former case,  $\chi=0.055$ in 
the latter case.}
\label{chivalues}
\end{figure}

\begin{figure}[ht]
\centering
\includegraphics*[width=1.\linewidth]{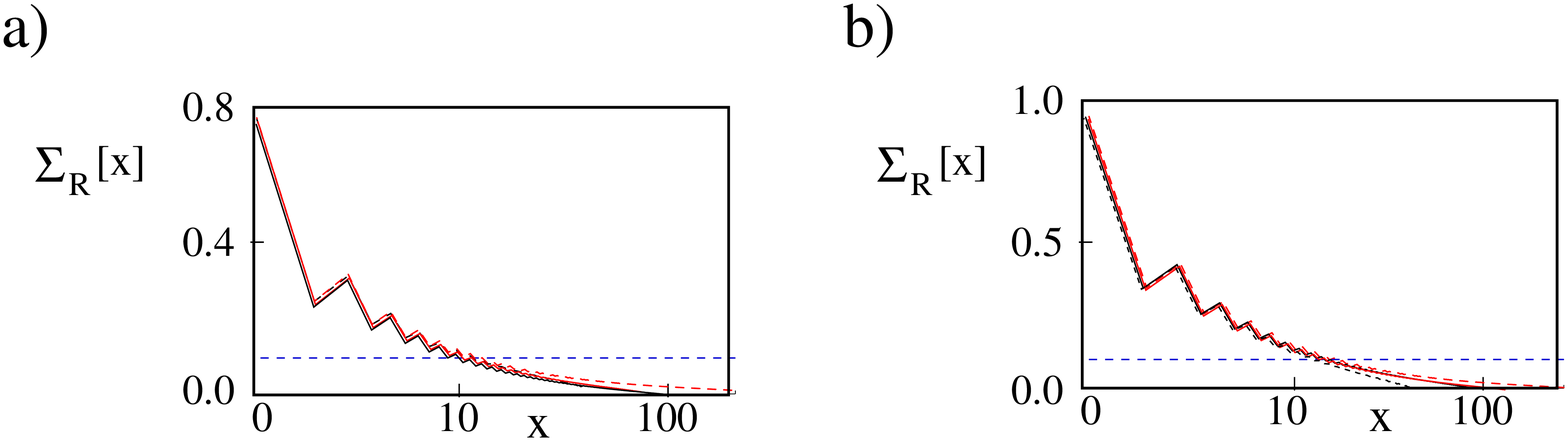}
\caption{\\
{\bf a):} Curves for ${\bf \Sigma}_R [ x ] $ at $\Delta = 0.3$,  $J_R^{'}/J = 0.6$, $J_L^{'} / J = 0.4$, and   $\ell = 50$  (dashed black curve),
$\ell = 100$ (solid red curve), $\ell = 150$ (solid black curve), $\ell = 300$  (dashed red curve). As a 
guide to the eye, the horizontal line at $y=0.1$ is shown as a dashed blue segment; \\
{\bf b):} Same as in panel {\bf a)}, but for $J_L^{'}/J=0.2$.  }
\label{sigmaR}
\end{figure}

\noindent
Next, we estimate the parameter $\chi$ in the three cases we consider. To do so, we just consider 
the value of the function  $ \tilde{\bf \Sigma} [ x ]= 
1  + \sum_{j = 1}^x \left\{ \frac{\langle S_{\bf G}^z S_{j , R}^z \rangle -\langle S_{\bf G}^z 
\rangle \langle S_{j , R}^z \rangle }{\langle (S_{\bf G}^z )^2 \rangle - ( \langle S_{\bf G}^z \rangle )^2 } \right\} $ at $x$ equal to the largest 
available value from simulations, $x = \ell = 300$. From the plots of $\tilde{\bf \Sigma} [ x ]$ reported in Fig.~\ref{chivalues}, we extract 
the values of $\chi$ reported in table~\ref{estimated}. At a given value of $J_L^{'} / J$, once $\chi$ is determined as 
discussed above, we extract $\xi_{K , R}$ by applying the KLTC to the function ${\bf \Sigma}_R [ x ]$ defined in Eq.~(\ref{n2c.6}) and plotted in Fig.\ref{sigmaR}. 

\begin{figure}[ht]
\centering
\includegraphics*[width=1.\linewidth]{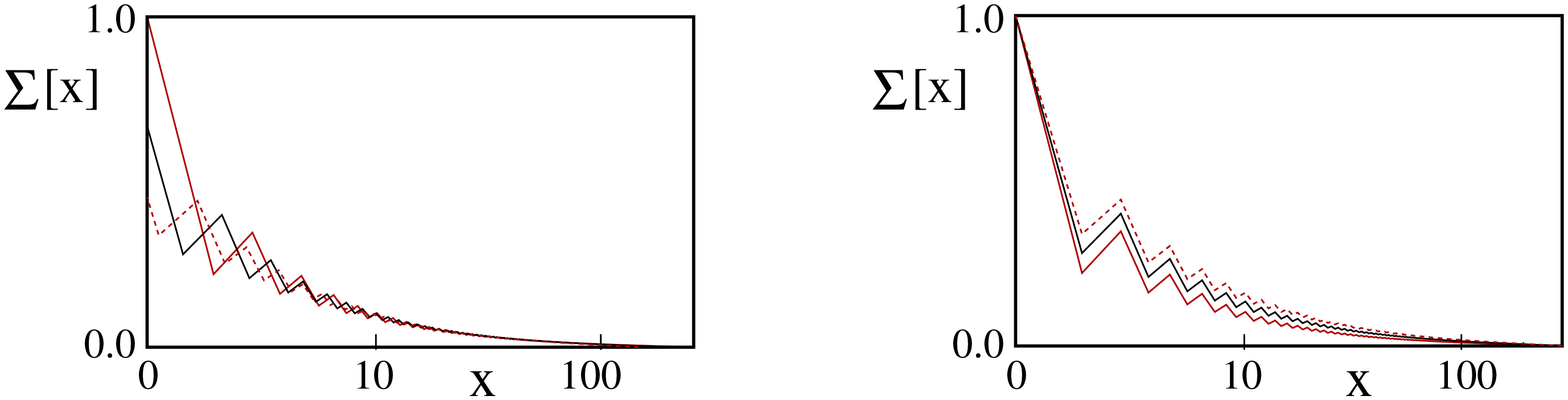}
\caption{\\
{\bf a):} Curves for ${\bf \Sigma} [ x ] $ at $\Delta = 0.3$, $\ell=300$, $J_R^{'}/J = 0.6$ and $J_L^{'} / J = 0.6$ (red dashed curve), 
$J_L^{'} / J = 0.4$ (black solid curve), and $J_L^{'} / J = 0.2$ (black solid curve): here the $x$ coordinate is rescaled 
with the ratio between the corresponding screening length, to induce curve collapse; \\
{\bf b):} Same as in panel {\bf a)}, but without rescaling $x$. }
\label{scalasym}
\end{figure}

The results, reported in the last column of table~\ref{estimated}, have 
an excellent consistency with the ones obtained 
for $\xi_K$ at the same values of $J_L^{'} / J$. This ultimately confirms our prediction  that the 
$\xi_K$ can be determined by assuming that the right-hand lead (the one feeling the stronger coupling to the 
impurity) screens an effective impurity larger by $1/2 - \chi$ than that would be at the symmetric point. In addition, 
we verify that, as expected, the residual weak-link interaction in Eq.~(\ref{mh.15}) does not induce any additional 
length scale associated with screening \cite{eggert99}. To do so, we resorted to SCT and plotted the curves 
for ${\bf \Sigma} [ x ]$ computed at $\Delta = 0.3$, $J'_R / J = 0.6$ and at $J^{'}_L /  J = 0.6,0.4,0.2$ 
and for $\ell = 300$ by 
rescaling the $x$ coordinate with the ratio between the corresponding $\xi_K$ and the $\xi_K$ computed at 
$J^{'}_L /  J = 0.6$ (result of table~\ref{estimated}). We plot the result in Fig.~\ref{scalasym}{\bf a)} 
where, for comparison, we also 
plot the same curves drawn without rescaling $x$  (Fig.~\ref{scalasym}{\bf b)}). Apparently, the 
excellent collapse in  Fig.~\ref{scalasym}{\bf a)}   evidences that no length scales but 
$\xi_K = \xi_{K , R}$ are dynamically generated by Kondo interaction.

\noindent

\noindent
 
\noindent
\vspace{0.5cm}
 
\begin{table}
\centering
\begin{tabular}{| c | c |c|c|}
 \hline 
 $J^{'}_L/J$ (at $J_R^{'} / J =0.6$) & Parameter $\chi$&$\xi_K$ (from ${\bf \Sigma} [ x ]$) & $\xi_{K , R}$ (from ${\bf \Sigma}_R [ x ]$)\\
 \hline
$0.6$ & $0.5$ & $10.23$ & $10.23$ \\
 \hline
$0.4$ & $0.238$ & $14.07$ & $13.87$ \\
 \hline
$0.2$ & $0.055$ & $18.13$ & $17.98$  \\
 \hline 
\end{tabular}
\caption{ Estimated values of the parameter $\chi$ (from Fig.~\ref{chivalues}), of $\xi_K$ (from Fig.~\ref{asym_1}) and 
of $\xi_{K , R}$ (from Fig.~\ref{sigmaR}) for $J'_{R} / J = 0.6$ and $J^{'}_L / J = 0.6,0.4,0.2$.}
\label{estimated}
\end{table}
\vspace{0.5cm}
 
To summarize, we may conclude that, for a magnetic impurity in an XXZ-chain, an $_L$-$_R$ asymmetry in the Kondo  
couplings does not spoil Kondo effect, as evidenced by scaling properties of the ${\bf \Sigma} [ x ]$.
However, it takes some important consequences in that it affects the distribution of the net screening 
between the leads, ultimately resulting in a renormalization of $\xi_K$ which, as a function of the weaker 
coupling, continuously evolves from the value it takes in the two-channel case (symmetric coupling), to 
the value it takes in the one-channel case (weaker coupling set to $0$) \cite{eggert92}. 
Conversely, on keeping $J_L^{'}$ fixed and {\it increasing} $J_R^{'}$, we expect a continuous shrinking 
of $\xi_K$. Eventually, when $J_R^{'} = J$, the right-hand lead plus ${\bf S}_{\bf G}$ turns into an 
uniform $\ell + 1$-site chain, coupled to the $\ell$-site left-hand lead via the weak-link Hamiltonian 
with parameters $J_L^{'} , J_{z , L}^{'}$. This suggests a first mean to experimentally trigger the 
crossover from Kondo effect to weak-link regime by continuously increasing $J_R^{'}$ till it becomes 
equal to $J$. At the same time, the expected continuous shrinking of $\xi_K$ makes it eventually 
become of the order of the  lattice site, which is appropriate when the onset of the weak link regime 
suppresses the scaling with $\xi_K$.

\subsection{Nonzero applied magnetic field}
\label{magimp}

We now discuss the effects of a nonzero $B$ at the impurity site, {\it i.e.,} the 
last term in Eq.~(\ref{modham.1}) on the Kondo screening and 
on the consequent value of $\xi_K$. In general, in the context of an XXZ 
spin chain, the effect of a finite uniform magnetic field in the $z$-direction can 
be accounted for by pertinently modifying the SLL approach \cite{hikifu_2}, which 
proves that a uniform field  only qualitatively affects Kondo effect at the impurity.
Also, when regarding the XXZ chain as an effective description of the Bose-Hubbard model, a uniform 
magnetic field arises from a uniform deviation of half-filling in the chemical potential of 
the Bose-Hubbard model which, again, does not 
qualitatively affect the system's behavior, at least as long as one works at 
finite particle number in the Bose-Hubbard 
model (canonical ensemble), corresponding to fixed $z$-component of the total spin in the 
XXZ chain \cite{grst_13,gst}. 

In the context of electronic Kondo effect, a nonzero $B$ has shown to result in a 
splitting in the Kondo resonance (with respect to the electron spin) that sets in 
at values of $B$ comparable with $T_K$. This comes together with a substantial 
suppression of the magnetoresistance/magnetoconductance across the Kondo impurity  
\cite{costi,otte}. As for what concerns the effects of a nonzero $B$ at an impurity
in a spin chain, to the best of our knowledge there is no, so far, a systematic study 
of how $B$  affects $\xi_K$ and, more in general, the development of the 
Kondo cloud. We now investigate this point by means of a combined use of the 
perturbative RG approach, based on the finite-$B$ RG equations in Eqs.~(\ref{rengu.2}),
and on DMRG approach to estimate $\xi_K$ at given values of the system parameters. 

\begin{figure}[ht]
\centering
\includegraphics*[width=.45\linewidth]{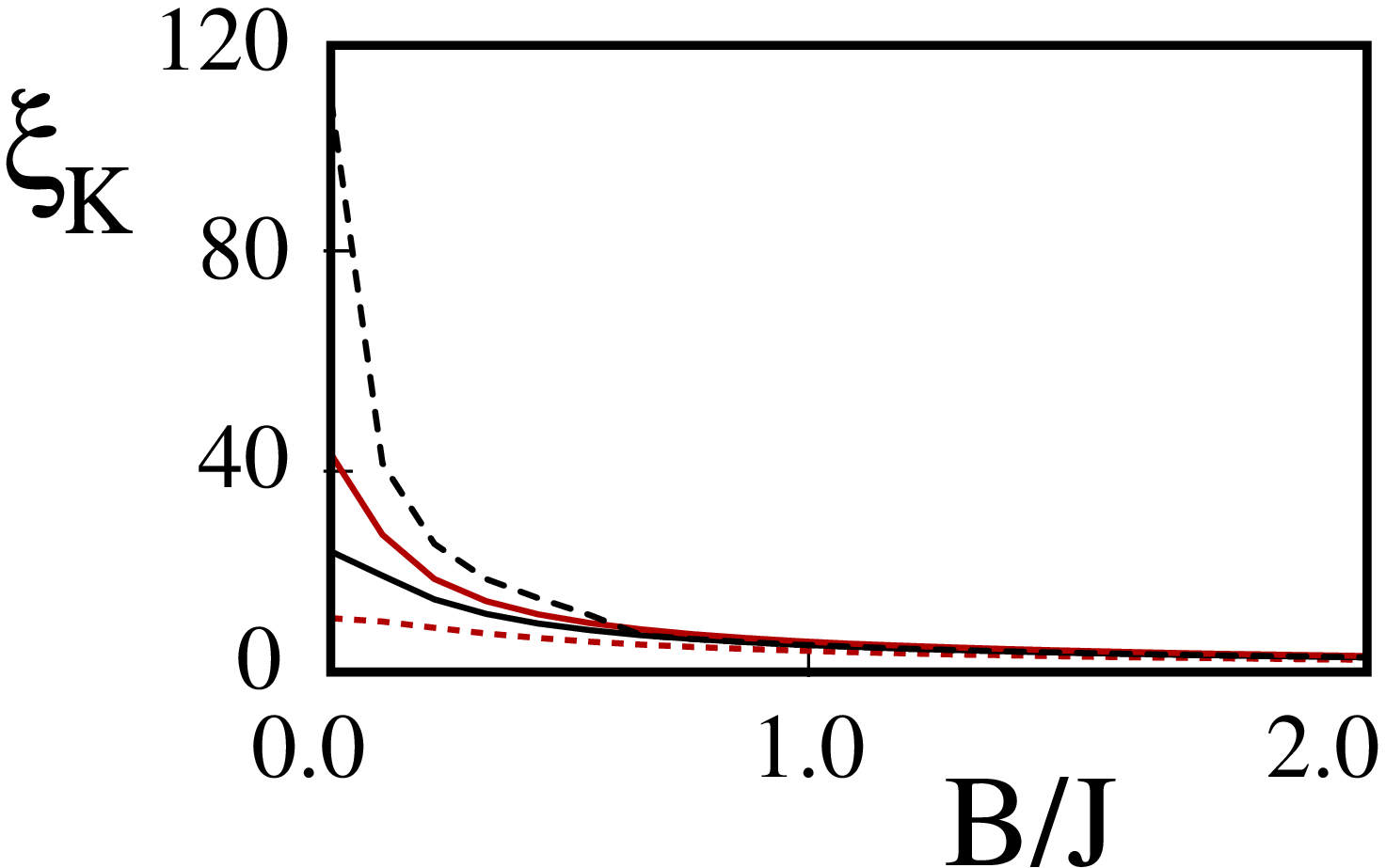}
\caption{$\xi_K [ J' / J , \Delta , B / J]$ {\it  vs.} $B/J$ derived from 
the integral curves corresponding to Eqs.~(\ref{rengu.2}) for $\Delta = 0.3$ and 
$J'/J=0.6$ (dashed red curve), $J' / J =0.4$ (solid black curve), $J' / J = 0.3$ (solid red curve), 
and $J' / J = 0.2$ (dashed black curve).}
\label{finite_b}
\end{figure}

\begin{figure}[ht]
\centering
\includegraphics*[width=.45\linewidth]{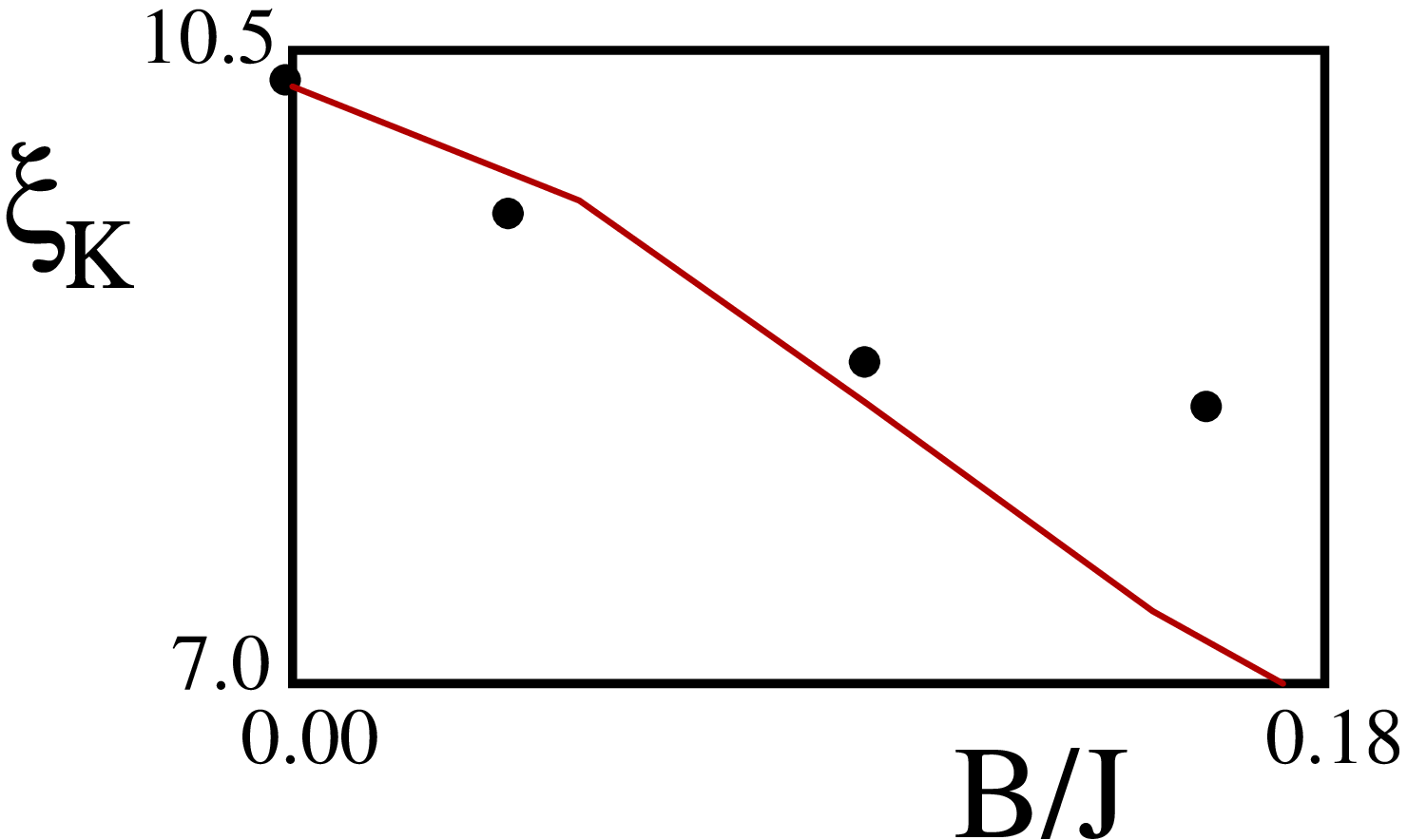}
\caption{\\ 
Solid red curve: 
$\xi_K [ J' / J , \Delta , B / J]$ {\it  vs.} $B/J$ derived for $J' / J = 0.6$ and 
$\Delta = 0.3$ from  the integral curves corresponding to Eqs.~(\ref{rengu.2}). \\ 
Black dots: $\xi_K [ J' / J , \Delta , B / J]$ {\it  vs.} $B/J$ derived for $J' / J = 0.6$ and 
$\Delta = 0.3$ by applying Kondo length collapse approach to the DMRG results obtained for 
$B/J=0.0,0.04,0.1,0.16$.}
\label{compar}
\end{figure}

Within perturbative RG approach, we integrate Eqs.~(\ref{rengu.2}) (which are 
expected to rigorously apply in the small-$B$ limit, that is, 
for $B / J \ll 1$), and use the integrated curves to define  a ``generalized'' Kondo length, $\xi_K [ J' /  J , \Delta , B / J ]$,
to be the scale at which the running couplings enter the nonperturbative regime, at 
given  $J', \Delta$ and $B$ (an important point to stress here is that, 
strictly speaking,  $\xi_K [ J' /  J , \Delta , B / J ]$ can be regarded as 
an actual Kondo length only as long as Kondo effect is not suppressed by 
$B$, that is, for $B < T_K$. At larger values of $B$, Kondo effect gets suppressed 
by Zeeman energy \cite{costi}, no Kondo length is dynamically generated though, still, 
$\xi_K [ J' /  J , \Delta , B / J ]$ keeps its meaning as  over-all length 
scale of the system). 
On numerically integrating Eqs.~(\ref{rengu.2}) we 
draw plots of $\xi_K [ J' /  J , \Delta , B / J ]$ {\it vs.} $B$ at 
fixed $J' / J$ and $\Delta$. In Fig.~\ref{finite_b} we plot 
$\xi_K [ J' /  J , \Delta , B /J ]$ {\it vs.} $B / J$ 
evaluated at $\Delta = 0.3$ and $J' / J =0.6,0.4,0.3,0.2$, after rescaling the 
values of $\xi_K$ taking into account the numerical value of the overall scale in the 
Kondo length evaluated with the KLCT. While the curves should actually be trusted only 
at small values of $B / J$, it is interesting to attempt to draw some qualitative 
conclusions by looking at a window of values of $B / J$ ranging from $0$ to $2$, which is 
what we do in Fig.~\ref{finite_b}. The main trend of all the plotted curves is a decrease  
in $\xi_K$ at small values of $B/J$ followed by a remarkable collapse of all the various 
$\xi_K$'s  onto a single value, of the order of 
a few lattice step, as $ B / J \sim 1$. To account for such a 
behavior, we observe that a nonzero $B$ 
introduces an additional ``magnetic'' length scale in the problem, $\xi_B = \alpha J / B$,
with $\alpha$ being a numerical factor of the order $1$, which we estimate later on from 
DMRG data. Accordingly, in employing the scaling approach, one has to properly modify   
Eq.~(\ref{icf.3}), consistently with what is done in 
Ref.~[\onlinecite{sholl}] for the Anderson impurity in an otherwise noninteracting 
electron chain. This eventually leads to a two-parameter scaling behavior,  that is,  
by denoting with ${\bf \Sigma}_B [ x ] $ the integrated spin correlation function 
at a nonzero $B$, we generalize Eq.~(\ref{icf.3}) as
\beq
{\bf \Sigma}_B [ x ] =  \sum_{ \bar{d}_b } \ell^{\bar{d}_b } \xi_{\bar{d}_b} \left[\frac{\xi_K}{\ell}   ; \frac{\xi_K}{\xi_B} ; \frac{x}{\xi_K} \right] 
\, , 
\label{lmf.1}
\eneq
\noindent
with, again, the sum taken over the scaling dimensions of the boundary operators entering 
the SLL representation for the XXZ spin chain with the local 
spin-1/2 impurity (note that in Eq.~(\ref{lmf.1}) we used $\bar{d}_b$ to denote a generic scaling dimension 
of a relevant boundary operator: using a different symbol from Eq.~(\ref{icf.3}) is motivated by the observation 
that, in principle, a nonzero $B$  breaks symmetries such as, for instance, spin-parity, thus potentially 
allowing the emergence of relevant boundary operators which were forbidden by symmetry at $B=0$).
To keep consistent with the zero-$B$ limit, as ``initial condition'' of Eq.~(\ref{lmf.1}) we require that 
\beq
\sum_{ \bar{d}_b } \ell^{\bar{d}_b } \xi_{\bar{d}_b} \left[\frac{\xi_K}{\ell}   ; 0 ; \frac{x}{\xi_K} \right] 
= \sum_{d_b } \ell^{d_b } \xi_{d_b} \left[\frac{\xi_K}{\ell}   ; \frac{x}{\xi_K} \right] = {\bf \Sigma} [ x ] 
\, . 
\label{lmf.2}
\eneq
\noindent 
From Eq.~(\ref{lmf.2}) we see that at small, but finite, values of $B / J$, ${\bf \Sigma} [ x ]$ is 
modified by a term $\propto B \xi_K$ with respect to its value at $B=0$, and so does 
$\xi_K$, as well. Moreover, a finite $B$ polarizes  ${\bf S}_{\bf G}$, so to break the 
${\bf S}_{\bf G}\to - {\bf S}_{\bf G}$-symmetry 
in the system Hamiltonian. The net average (``static'') polarization of ${\bf S}_{\bf G}$ corresponds to 
a reduction in the fluctuation of the local impurity spin. Since the finite extension of $\xi_K$ is 
a consequence of the dynamical mechanism of Kondo screening (related to the fluctuations in 
${\bf S}_{\bf G}$), the smaller the fluctuations are, the less spins are needed to dynamically
screen the impurity spin. Therefore, one naturally expects that a nonzero $B$ implies a 
reduction in $\xi_K$, as it appears from the plots in Fig.~\ref{finite_b}. This can be ultimately 
inferred from Eq.~(\ref{lmf.2}) taken in the limit $\xi_K , \xi_B , x \ll \ell$, required to 
suppress finite-size corrections to scaling, and $\xi_B \gg \xi_K$, corresponding to small 
values of $B$. In this limit, $\xi_K$ works as a reference length scale, and 
Eq.~(\ref{lmf.1}) simplifies into 
\beq
{\bf \Sigma}_B [ x ]  \approx \hat{\bf \Sigma}_{1 , B}[ x ] =   \sum_{ \bar{d}_b } \ell^{\bar{d}_b }
\xi_{\bar{d}_b} \left[ 0  ; \frac{\xi_B}{\xi_K} ; \frac{x}{\xi_K} \right] 
\, . 
\label{lmf.2x}
\eneq
\noindent
Eq.~(\ref{lmf.2x}) determines $\xi_K [ J' /  J , \Delta , B /J ]$ from the KLCT condition
\beq
  \sum_{ \bar{d}_b } \ell^{\bar{d}_b } \xi_{\bar{d}_b} \left[ 0  ; \frac{\xi_B}{\xi_K [ J' /  J , \Delta ] } ; 
  \frac{\xi_K [ J' /  J , \Delta , B /J ]}{\xi_K [ J' /  J , \Delta  ] } \right] 
  = r
  \, ,
\label{lmf.x1}
\eneq
\noindent
where we choose $r=0.1$. Increasing $B$ from $B$ to $B' = \rho B$ ($\rho > 1$), we therefore obtain 
\beq
  \sum_{ \bar{d}_b } \ell^{\bar{d}_b } \xi_{\bar{d}_b} \left[ 0  ; \frac{\xi_{B' }}{\rho \xi_K [ J' /  J , \Delta ]     } ; \frac{\rho \xi_K [ J' /  J , \Delta , B' /J ]   }{
  \rho \xi_K [ J' /  J , \Delta ]   } \right] =\sum_{ \bar{d}_b } \ell^{\bar{d}_b } \xi_{\bar{d}_b} \left[ 0  ; \frac{\xi_B}{\xi_K [ J' /  J , \Delta ] } ;
  \frac{\xi_K [ J' /  J , \Delta , B /J ]}{\xi_K [ J' /  J , \Delta ]  } \right]  
  \, ,
\label{lmf.x2}
\eneq
\noindent
which implies a reduction of $\xi_K [ J' /  J , \Delta , B' /J ] $ by a factor $\rho^{-1} = B / B'$. 
To check this conclusion, we compare the values of $\xi_K [ J' / J , \Delta , B / J ]$ obtained from the integrated 
Eqs.~(\ref{rengu.2}) at $J' / J = 0.6$ and $\Delta = 0.3$ as a function of $B  / J$ with the estimates we derive by 
applying the KLCT  to the DMRG results at the same values of $J' / J$ and $\Delta$ and 
at the selected values of $B / J$. Note that, in order to enhance the window of 
expected validity of Eqs.~(\ref{rengu.2}), 
we have chosen the largest possible value of $J' / J$ among the ones we consider in this work, so to 
minimize the corresponding value of $\xi_K$). In Fig.~\ref{compar}, we plot 
$\xi_K [ 0.6  , 0.3 , B / J ]$ 
{\it vs.} $B / J$ derived from Eqs.~(\ref{rengu.2}) for $0 \leq B / J \leq 0.18$, 
and we display as black dots the values of 
$\xi_K [ 0.6  , 0.3 , B / J ]$ estimated within the Kondo length collapse technique applied to the DMRG results
for $B / J = 0.0, 0.04,0.1,0.16$. We see that the dots lie quite close to the curve for $B / J \leq 0.1$, so, we
infer a validity of the analytical RG Eqs.~(\ref{rengu.2}) for values of $B$ less or equal to 
ten percent of the high-energy cutoff ($\sim J$). Beyond those values of $B / J$, we may extrapolate that the 
DMRG results are systematically larger than the predictions of the perturbative RG 
approach, which is consistent with the fact that the latter technique systematically underestimates 
higher-order fluctuations, that are ultimately responsible for the size of the Kondo cloud 
\cite{zawad_1,zawad_2,zawad_3}.
 
\noindent

\noindent

\begin{figure}[ht]
\centering
\includegraphics*[width=.45\linewidth]{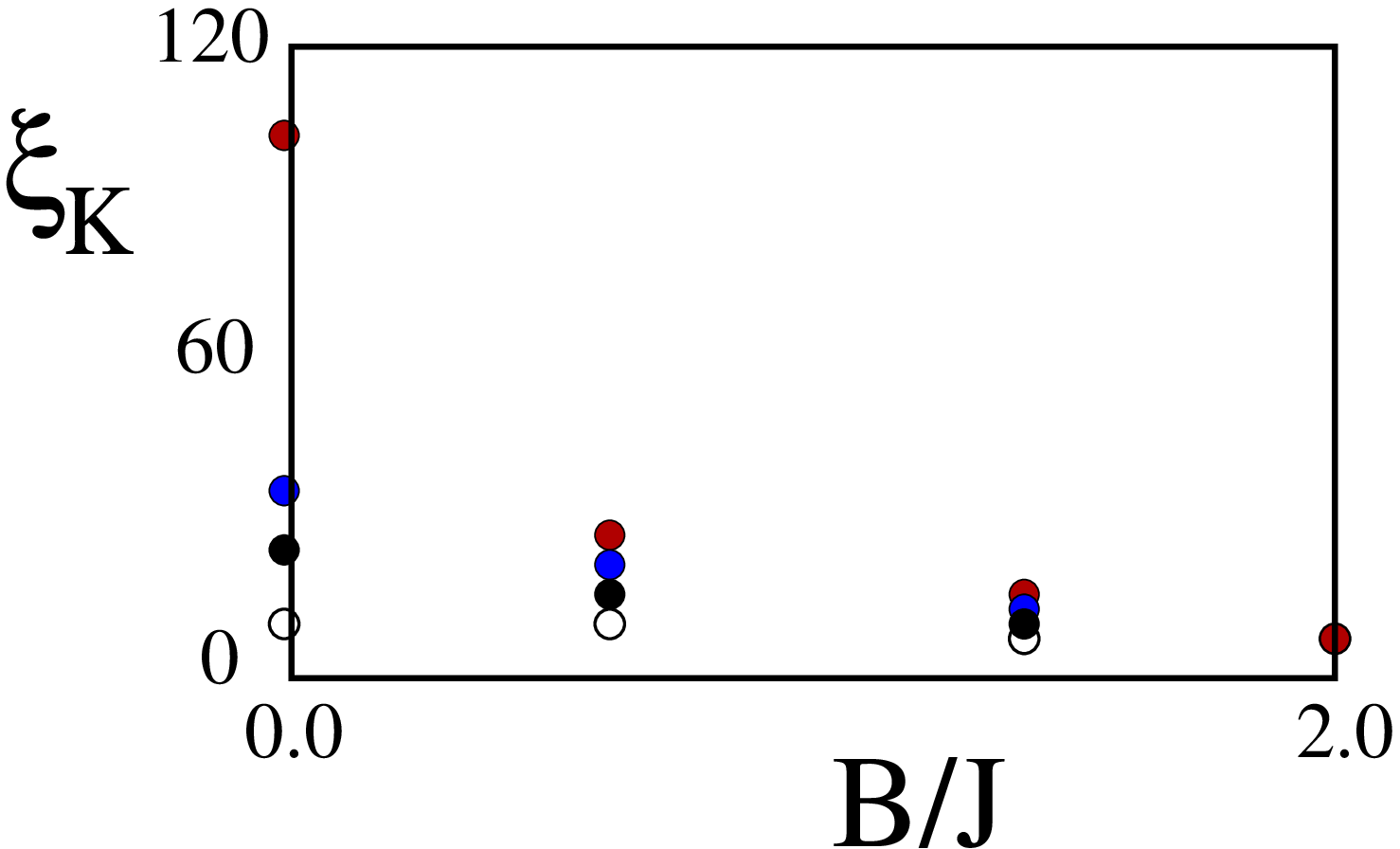}
\caption{$\xi_K [ J' / J , \Delta , B / J]$ {\it  vs.} $B/J$ derived for 
$\Delta = 0.3$  by applying KLCT  to the DMRG results obtained for 
$J' /J = 0.2$ (full red dots), $J'/J = 0.4$ (full blue dots), $J' / J = 0.3$ (full black dots), $J' / J = 0.6$ (empty 
dots), at $B / J = 0.0,0.6,1.4,2.0$. }
\label{pointplot}
\end{figure}

The second remarkable feature shown in Fig.~\ref{finite_b} is that increasing $B$ all the $\xi_K$'s 
collapse onto a single value, of the order of a few lattice steps.
To understand this, we note that, as $B / J \sim 1$ and, accordingly, $\xi_B \ll \xi_K$, Eq.~(\ref{lmf.1}) 
simplifies  into 
\beq
{\bf \Sigma}_B [ x ]  \approx \hat{\bf \Sigma}_{2 , B}[ x ] =    \lim_{ y \to \infty} \sum_{ \bar{d}_b } 
\ell^{\bar{d}_b } \xi_{\bar{d}_b} \left[ 0  ; y  ; \frac{x}{\xi_B} \right] 
\, .
\label{lmf.3}
\eneq
\noindent
Eq.~(\ref{lmf.3}) again displays an universal scaling function, but now scaling with $\xi_B$ being 
the reference length scale, since any reference to the value of $J' / J$ has disappeared. This  
eventually accounts for the collapse of all the Kondo lengths onto a $J'$-independent 
value, at large enough values of $B$. To confirm this result with our numerical analysis, 
we applied KLCT to DMRG data for ${\bf \Sigma} [ x ]$ derived at $\ell = 300$ for   $J' /  J = 0.2,0.3,0.4,0.6$ at 
$ B / J = 0.0,0.6,1.4,2.0$. We plot our result in Fig.~\ref{pointplot}, from which we see that, as soon 
as $B / J$ takes off,  a remarkable collapse of the scaling lengths at various 
values of $J' / J$ sets in (note that this also allows us to estimate $\alpha \approx 3.5$). 
In Fig.~\ref{pointplot} we ultimately see an evident large-$B$ collapse, as 
predicted by Eqs.~(\ref{rengu.2}), though up to an over-all numerical factor.
 
\noindent

To conclude this section, a comment is in order about the possibility of 
using $B$ as a control parameter to drive the system along a 
crossover from a Kondo-like behavior to a weak-link like behavior. 
We note that, in the $B / J \gg 1$-limit, one of 
the two impurity levels is pushed very high in energy, with respect to 
the other one. This strongly suppresses processes in which ${\bf S}_{\bf G}$ switches 
between the two eigenstates of $S_{\bf G}^z$, leaving them only as virtual 
processes. To take this into account, one may resort to an effective, low energy 
description of the impurity dynamics. Summing over virtual processes leads 
to a second-order (in the $J'$'s) weak-link Hamiltonian, of the form 
\beq
H_B \sim - \frac{J_L^{'} J_R^{'} }{2 | B | } \{ S_{1 , L }^+ S_{1 , R}^- + S_{1 , L }^- S_{1 , R}^+ \}
+ \ldots \, , 
\label{bf.1}
\eneq
\noindent
with the ellipses standing for subleading corrections to $H_B$. $H_B$ in Eq.~(\ref{bf.1}) corresponds to 
a weak-link Hamiltonian which is expected to behave, under scaling, completely differently from 
a Kondo-like Hamiltonian. 

Thus, we see that increasing $B$ works as an alternative 
(to acting on channel anisotropy) knob to tune the crossover from 
Kondo effect to weak-link regime. While 
it is qualitatively analogous to increasing the couplings of 
${\bf S}_{\bf G}$ to one lead keeping the 
other fixed, it is definitely different with respect 
to possible experimental realizations of either method.
Indeed, tuning $B$ means acting on a single lattice sites. At variance, 
acting onto one of the two bond impurities leaving the other unaltered 
implies pertinently adjusting a single-bond coupling strength. Both operations 
can be in principle implemented in e.g. cold-atom realization of the XXZ spin chain
and one can choose either one, according to which one is easier to operate. More specifically, 
using the notations of section~\ref{moha}, for $B=0$ (no added on-site potentials) 
having $\sigma \lesssim d$ one can fix the added right potential intensity $V_R$ (which fixes $J_R^{'}$), 
and vary the left one, $V_L$. For $V_L>V_R$ one has $J_L^{'}<J_{R}^{'}$ and for 
$V_L \gg V_R$, one has $J_L^{'} \ll J_R^{'}$ and the one-channel Kondo physics 
is retrieved, which is the case studied in 
section~\ref{asym}. When at variance $V_L<V_R$, then $J_L^{'}>J_{R}^{'}$, 
and for $V_L=0$ then only one link -- in the middle of the chain -- 
is altered and $J_L^{'}$ is equal to the bulk value, and the physics of the 
weak-link is retrieved. In practice, we expect that the Kondo length decreases 
from its value at $J_L^{'}=0$ (for $V_L \gg V_R$) to smaller values, arriving 
to be order of the lattice spacing for $V_L \ll V_R$. It would be interesting 
as a future study to quantitatively analyze this crossover at $B=0$ 
from the Kondo to the weak-link regime, which we expect to be similar 
to the crossover studied increasing $B$ in the present section.

\section{Conclusions}
\label{concl}

By combining the renormalization group approach with the numerical density-matrix renormalization group 
technique, we have studied in detail the Kondo screening length at a magnetic impurity in the middle of a spin-1/2 
XXZ spin chain. The combination of the two methods allowed us to exactly derive 
the dependence of $\xi_K$ on the various system parameters, as well as to provide 
a systematic physical interpretation of its behavior when, for instance, the magnetic impurity is 
separately coupled to two different leads, and/or a nonzero magnetic field applied to 
the impurity induces a crossover from a Kondo impurity to a weak-link. 
To this aim, we have generalized to spin-Kondo effect in the XXZ chain the 
method of extracting $\xi_K$ from the scaling properties of the integrated 
real-space spin correlation functions, used in Refs.~[\onlinecite{baraf,sholl}] 
for ``conventional'' Kondo effect in metals. 

Our technique enabled us to provide realistic estimates for $\xi_K$ from 
$10$ to $100$ lattice sites, to systematically discuss how it varies as a function 
of the asymmetry in the couplings to the two channels and, eventually, to 
map out the shrinking of the Kondo length that characterizes the crossover from
Kondo impurity- to weak-link-physics in the presence of a large value of 
the magnetic field $B$ applied to the impurity. As real-space equal-time spin correlations are 
measurable, {\em e.g.,} in ultracold realizations of the homogeneous XXZ spin chain \cite{gst}, 
we believe that our results suggest a 
new way to measure the (so far) rather elusive Kondo screening length. We observe that 
in metals the Kondo length is expected to be of the order of  thousands of the lattice spacings, but the overlap 
of different Kondo cloud makes it difficult to detect the Kondo length. In quantum spin chains we get 
typical values of $\sim 10-100$ lattice spacings, which is realistic for experimental implementation 
of the XXZ chain with ultracold atoms, and at the same time tunable (unlike what happens in metals) varying 
the ratio $J'/J$.  Moreover, we note that our derivation is based on properties of quantities, such as 
real-space spin-spin correlation functions, which can be experimentally accessed by 
measuring the density-density correlations and their spatial integrals,
as discussed, for instance, in Refs.[\onlinecite{bloch,greiner}]. 
Therefore, we see that directly accessing in a realistic experiment the real-space correlation functions
at low enough temperatures and, therefore, probing the Kondo screening length is already a possibility  
within the reach of nowadays technology.

In view of the fact that both the XXZ spin chain and its spin-liquid phase and 
a 1D system of spinless interacting electrons are described as 
a spinless Luttinger liquid with suitably chosen parameters, our approach can 
be straightforwardly generalized to Kondo effect in the presence of interacting 
electronic leads \cite{furunaga,joha_1,joha_2}. 

Finally, we observe that in the paper we considered (one or) two tunable bond impurities. 
However, in experimental implementations of the XXZ model for ultracold atoms in optical lattices 
one may think to alter the couplings ({\it i.e.}, the tunnelings) by using localized 
external potentials via laser beams with width $\sigma$ applied on the quantum gas. The fact that one cannot perfectly 
center these additional potentials exactly between two lattice sites finally results in an asymmetry 
of the couplings (see Ref.~[\onlinecite{gst}]), such as the one we discuss here. However, generically the added potentials 
will have a width $\sigma$ larger than one or two lattice sites spacing. Therefore, one has 
to consider an extended region of width $\sigma$ in which the couplings are altered. 
Since the Hamiltonian ${\cal H}$ in Eq.~(\ref{modham.1}) 
can also be regarded as an effective description of an extended spin cluster coupled to two homogeneous 
XXZ chains (as we outline in appendix~\ref{mappings}), it would be interesting to generalize 
our combined approach to study the crossover from  Kondo effect to  weak-link regime  in the case of 
a finite central region of ``realistic'' shapes, such 
as Gaussian. Of course, this requires going through a number of subtleties, both on 
the formal/analytical side as well as on the numerical side, on how to define and extract 
the Kondo length for these extended multi-bond impurities.
Yet, this line of work is important both to understand the persistence of the Kondo 
effect for extended defects and to address 
the applicability of our model to realistic systems, and we plan to 
leave this as the subject of future investigations.

\vspace{0.5cm}

{\bf Acknowledgements --} 

We thank I. Affleck, R. Pereira and P. Sodano for valuable discussions.

\appendix

\section{Mapping of extended many-spin regions onto effective weak-link and single-impurity Hamiltonians}
\label{mappings}

In this appendix we show, by means of a few paradigmatic examples, that ${\cal H}$ in Eq.~(\ref{modham.1}) can be  
regarded as an effective description of a generic $M$-spin extended region ${\cal R}$ in the middle 
of the chain, weakly coupled to the rest of the chain through its endpoints. To do so, 
we start with the reference Hamiltonian  $H$  given by 
\beq
H  = \sum_{X = L , R } H_X + H_{\cal R} + H' \, , 
\label{map.1}
\eneq
\noindent
with $H_L$ and $H_R$ as in Eq.~(\ref{modham.1}), and 
\begin{eqnarray}
H_{\cal R} &=& \bar{J} \sum_{ j = 1}^{M - 1 } \: \{ S_{j , {\cal R} }^+ S_{ j + 1 , {\cal R}}^- + 
S_{ j , {\cal R}}^- S_{j + 1 , {\cal R}}^- + \Delta S_{j , {\cal R}}^z S_{ j + 1 , {\cal R}}^z \} +  \bar{B} \sum_{ j = 1}^{M  } \:   S_{j , {\cal R} }^z 
\nonumber \\
H' &=& J'_L \: \{ S_{ 1 , L}^+ S_{1 , {\cal R}}^+ + S_{1 , L}^- S_{1 , {\cal R}}^+  \} + J^{'}_{z ,L}  S_{ 1 , L }^z S_{ 1 , {\cal R}}^z  
+ J'_R \: \{ S_{ 1 , R}^+ S_{M , {\cal R}}^+ + S_{1 , R}^- S_{M , {\cal R}}^+  \} + J^{'}_{z ,R}  S_{ 1 , R }^z S_{ M , {\cal R}}^z 
\, . 
\label{map.2}
\end{eqnarray}
\noindent
Note that, in Eq.~(\ref{map.2}), for the sake of simplicity we have set the spin exchange strengths, 
as well as the applied magnetic field $\bar{B}$, to be uniform in the $H_{\cal R}$-term. 
Yet, we expect no particular complications to 
arise in the more general case of nonuniform parameters. As $\bar{B} = 0$, symmetry of $H_{\cal R}$ under spin-parity, 
${\bf S}_{ j , {\cal R}} \longrightarrow - {\bf S}_{ j , {\cal R}}$, implies that its ground state is 
either twofold degenerate, or nondegenerate, according to whether $M$ is odd, or even. This 
suggests to make, in the following, two separate discussions for the odd-$M$ and for the even-$M$ case, respectively.

\subsection{Mapping for $M=3$}
\label{mapspin}

Besides the trivial case $M=1$, $M=3$ corresponds to the prototypical situation in 
which $H_{\cal R}$ is expected to map onto an effective spin-1/2 Kondo 
impurity Hamiltonian. To illustrate how this works, let us start by assuming $\bar{B} = 0$. 
For $M=3$, we therefore obtain  
\beq
H_{\cal R} = \bar{J} \{ S_{1 , {\cal R}}^+ S_{2 , {\cal R}}^- + S_{2 , {\cal R}}^+ S_{3 , {\cal R}}^- + {\rm h.c.} \} 
+ \bar{J} \Delta \{ S_{ 1 , {\cal R}}^z + S_{3 , {\cal R}}^z \} S_{2 , {\cal R}}^z
\, . 
\label{maps.1}
\eneq
\noindent
The ``natural'' basis for the Hilbert space of $H_{\cal R}$ is the one made out of 
the simultaneous eigenstates of $ S_{ 1 , {\cal R}}^z , S_{2 , {\cal R}}^z , S_{3 , {\cal R}}^z$, which we label as
$ | \sigma_{1} , \sigma_2 , \sigma_3 \rangle$. Making $H_{\cal R}$ act onto each one of 
the $8$ basis states above, we obtain 
\begin{eqnarray}
 &&H_{\cal R} | \! \uparrow , \uparrow , \uparrow \rangle  = \frac{\bar{J} \Delta}{2}  | \! \uparrow , \uparrow , \uparrow \rangle \nonumber \\
  &&H_{\cal R} | \! \downarrow , \downarrow , \downarrow \rangle  = \frac{\bar{J} \Delta}{2}  | \! \downarrow , \downarrow , \downarrow \rangle \nonumber \\
 && H_{\cal R} | \! \uparrow , \uparrow , \downarrow \rangle = \bar{J} | \! \uparrow , \downarrow , \uparrow \rangle \nonumber \\
 && H_{\cal R} | \! \uparrow , \downarrow , \uparrow \rangle = \bar{J} \{ | \! \uparrow , \uparrow , \downarrow \rangle 
 + | \! \downarrow , \uparrow , \uparrow \rangle \} - \frac{\bar{J} \Delta}{2} | \! \uparrow , \downarrow , \uparrow \rangle \nonumber \\
  && H_{\cal R} | \! \downarrow , \uparrow ,\uparrow \rangle = \bar{J} | \! \uparrow , \downarrow , \uparrow \rangle \nonumber \\
  && H_{\cal R} | \! \downarrow ,  \downarrow , \uparrow \rangle = \bar{J} | \! \downarrow , \uparrow , \downarrow   \rangle \nonumber \\
 && H_{\cal R} | \! \downarrow ,  \uparrow , \downarrow  \rangle = \bar{J} \{ | \! \downarrow ,  \downarrow , \uparrow  \rangle 
 + | \! \uparrow , \downarrow , \downarrow \rangle \} - \frac{\bar{J} \Delta}{2} | \! \downarrow ,  \uparrow , \downarrow  \rangle \nonumber \\
   && H_{\cal R} |\uparrow ,  \downarrow ,  \downarrow    \rangle = \bar{J} | \! \downarrow , \uparrow , \downarrow   \rangle 
 \, .
 \label{maps.2}
\end{eqnarray}
\noindent
As a result,  the lowest-energy eigenvalue of 
$H_{\cal R}$ is $\epsilon = - \frac{\bar{J} \Delta}{4} - \sqrt{2 \bar{J}^2 + 
\left( \bar{J} \Delta / 4 \right)^2}$. As expected, for $\bar{B}=0$ $\epsilon$ is twofold degenerate, 
with corresponding eigenstates given by 
\begin{eqnarray}
 | \! \Uparrow \rangle &=& \frac{1}{2 \sqrt{\cosh ( \xi ) } }\{  e^{ - \frac{\xi}{2} } [ | \! \uparrow , \uparrow , \downarrow \rangle 
 +  | \! \downarrow , \uparrow , \uparrow   \rangle ]
 - \sqrt{2} e^\frac{\xi}{2} | \! \uparrow , \downarrow , \uparrow \rangle \} \nonumber \\
  | \! \Downarrow \rangle &=&\frac{1}{2 \sqrt{\cosh ( \xi ) } } \{  e^{ - \frac{\xi}{2} } [
  | \! \downarrow , \downarrow , \uparrow \rangle +  | \! \uparrow , \downarrow , \downarrow   \rangle ] 
 - \sqrt{2} e^\frac{\xi}{2} | \! \downarrow , \uparrow , \downarrow \rangle \} \, , 
 \label{maps.3}
\end{eqnarray}
\noindent
with 
\begin{equation}
 \cosh ( \xi ) = \frac{\sqrt{2 \bar{J}^2 + 
\left( \frac{\bar{J} \Delta}{4} \right)^2}}{\sqrt{2} \bar{J}} \,, \qquad
 \sinh ( \xi ) = \frac{\left( \frac{\bar{J} \Delta}{4} \right)}{\sqrt{2} \bar{J}} \; . 
 \label{maps.4}
\end{equation}
\noindent
To complete the mapping onto an effective spin-1/2 Kondo Hamiltonian, we need to resort to 
an effective low-energy formulation of the dynamics of ${\cal R}$ only involving 
the states in Eqs.~(\ref{maps.3}). To do so, we employ the projection operator over the corresponding 
subspace of the Hilbert space, 
${\cal P}_{\cal R} = \sum_{\rho = \Uparrow , \Downarrow} \: | \rho \rangle \langle \rho |$. Within the
low-energy subspace of the Hilbert space, we also  define the ``collective'' spin-1/2
operators for the central region, $S_{\bf G}^a$, as 
\begin{equation}
 S_{\bf G}^+ \equiv - | \! \Uparrow \rangle \langle \Downarrow \! | \,, \qquad
  S_{\bf G}^- \equiv - | \! \Downarrow \rangle \langle \Uparrow \! | \,, \qquad
  S_{\bf G}^z \equiv \frac{1}{2} \sum_\rho \: \rho | \rho \rangle \langle \rho | \; . 
\label{maps.5}
\end{equation}
\noindent
Now, by direct calculation, one finds  
\begin{eqnarray}
 && \langle \Downarrow \! | S_{1 , {\cal R}}^- | \! \Uparrow \rangle = \langle \Downarrow \! | S_{ 3 , {\cal R}}^-
 | \! \Uparrow \rangle = - \frac{1}{\sqrt{2} \cosh ( \xi)}, \nonumber \\
  && \langle \Uparrow \! | S_{ 1 , {\cal R}}^+ | \! \Downarrow \rangle = \langle \Uparrow \! | S_{ 3 , {\cal R}}^+ 
  | \! \Downarrow \rangle = - \frac{1}{\sqrt{2} \cosh ( \xi)}, \nonumber \\
   && \langle \rho | S_{ 1 , {\cal R}}^z | \rho \rangle =   \langle \rho | S_{ 3 , {\cal R}}^z | \rho \rangle =  \frac{e^\xi}{2 \cosh ( \xi)} 
   \, , 
   \label{maps.6}
\end{eqnarray}
\noindent
all the other matrix elements being equal to $0$. As a result, we obtain 
\beq
{\cal P}_{\cal R} H' {\cal P}_{\cal R} = \tilde{J}' \: \{ [ S_{1 , L}^+ + S_{1 , R}^+ ] S_{\bf G}^- + 
[ S_{1 , L}^-  + S_{1 , R}^- ] S_{\bf G}^+ \} + \tilde{J}_z^{'}  [ S_{1 , L}^z + S_{1 , R}^z ] S_{\bf G}^z
\, , 
\label{maps.7}
\eneq
\noindent
with 
\begin{equation}
 \tilde{J}' =  \frac{J'}{\sqrt{2} \cosh ( \xi)} \,, \qquad
 \tilde{J}_z^{'} = \frac{e^\xi \Delta J'}{2 \cosh ( \xi) } \, . 
 \label{maps.8}
\end{equation}
Eq.~(\ref{maps.7}) takes the desired form of an effective spin-1/2 impurity Kondo Hamiltonian. 
Note that, at variance with the case $M=1$, in using the Hamiltonian in Eq.~(\ref{maps.7}) to 
perform the perturbative RG analysis, one has to cut off the dynamics to energy scale 
of the order of the energy gap between the states $ | \rho \rangle$ and 
the next excited eigenstates of $H_{\cal R}$. This implies a ``cutoff renormalization'', 
from $D_0 \sim J $ to $D_0 \sim \bar{J}$, potentially leading to 
an unavoidable renormalization to lower values of the Kondo temperature and, correspondingly, to 
higher values of $\xi_K$. Clearly, a finite $\bar{B}$ breaks the twofold ground-state degeneracy of 
$H_{\cal R}$, resulting into an additional term $B S_{\bf G}^z$ to add at the right-hand side of 
Eq.~(\ref{maps.7}), with $B \propto \bar{B}$.

\subsection{Mapping for $M=2$}
\label{wlink}

The simplest even-$M$ central region is realized as a single weak-link, corresponding 
to $M=0$, in which case one obtains 
\beq
H_{\cal R} = J' \{ S_{1 , L}^+ S_{1 , R}^- + S_{1 , L}^- S_{1 , R}^+ \} + J_z^{'} S_{1 , L}^z S_{1 , R}^z 
\, . 
\label{wlink.0}
\eneq
\noindent
Besides $M=0$, the first nontrivial 
case corresponds to $M=2$, that we discuss in the following. Again, for 
the sake of simplicity, we start our analysis by assuming $\bar{B} =0$. In this case, 
the ground state of ${\cal R}$ is nondegenerate. To construct it, we start by recovering 
the action of  $H_{\cal R}$ on the set of the simultaneous eigenstates of 
$S_{1 , {\cal R}}^z $ and $S_{2 , {\cal R}}^z$,  $ | \sigma_1 , \sigma_2 \rangle$. Specifically, we obtain 
\begin{eqnarray}
&& H_{\cal R}| \uparrow , \uparrow \rangle = \frac{ \bar{J} \Delta}{4} 
\: | \uparrow , \uparrow \rangle, \nonumber \\
&& H_{\cal R}| \downarrow , \downarrow \rangle = \frac{ \bar{J} \Delta}{4}
\: | \downarrow , \downarrow \rangle, \nonumber \\
&& H_{\cal R} \frac{1}{\sqrt{2}} \{ | \uparrow , \downarrow \rangle \pm | \downarrow , \uparrow \rangle \}
= \left\{ \pm \bar{J} - \frac{ \bar{J} \Delta}{4} \right\} \:   
\frac{1}{\sqrt{2}} \{ | \uparrow , \downarrow \rangle \pm | \downarrow , \uparrow \rangle \}
\, . 
\label{wlink.1}
\end{eqnarray}
\noindent
From Eqs.~(\ref{wlink.1}), we find that the ground state of $H_{\cal R}$, 
$ | {\bf 0}\rangle$, is the nondegenerate singlet $ | {\bf 0} \rangle = 
\frac{1}{\sqrt{2}} \{ | \uparrow , \downarrow \rangle - | \downarrow , \uparrow \rangle \}$.
Let ${\cal P}_{\bf 0}$ be the projector onto $ | {\bf 0}\rangle$. We obtain 
\beq
{\cal P}_{\bf 0} H' {\cal P}_{\bf 0} = 0 \, ,
\label{wlink.2}
\eneq
\noindent
which implies that the first nontrivial contribution to the effective weak-link 
Hamiltonian arises to second-order in $J'$. This is recovered within 
a systematic Schrieffer-Wolff (SW) procedure, eventually yielding the effective weak 
link Hamiltonian $H_{\rm WL}^{\rm Eff}$ \cite{swolff} given by 
\beq
H_{\rm WL}^{\rm Eff} = \frac{( J' )^2 }{2 \bar{J} \left( 1 + \frac{\Delta}{2} \right)}
 \: \{ S_{1 , L}^+ S_{1 , R}^- + S_{1 , L}^- S_{1 , R}^+ \} + 
 \frac{( J' \Delta )^2 }{4  \bar{J} } S_{1 , L}^z S_{1 , R}^z
 \, . 
 \label{wlink.3}
 \eneq
\noindent
Eq.~(\ref{wlink.3}) ultimately shows that an $M=2$ central region (and, more generally, an even-$M$
central region with $\bar{B} = 0$), can be regarded as a simple weak-link, 
at least as long as the involved energies are lower than  
the energy gap between $ | {\bf 0} \rangle$ and the first 
excited eigenstate(s) of $H_{\cal R}$. As $ | {\bf 0}\rangle$ is a spin singlet, a non vanishing $\bar{B}$ 
does not alter 
this picture, at least as long as $\bar{B} \ll J$. Remarkably, as a finite $\bar{B}$ breaks the ground-state twofold degeneracy 
for $M$ odd, it can make the junction with $M=3$ effectively behave as a weak-link, as well. To spell this 
out, let us set $M=3$ and assume $B>J$. Let us set 
${\cal P}_-$ to be the projector onto the eigenstate of $S_{\bf G}^z$ belonging to 
the eigenvalue $-1/2$. To leading order in the boundary couplings, one may again employ 
the SW procedure, to resort to a ${\cal P}_-$ projected effective 
Hamiltonian for the whole chain. The result is 
\begin{equation}
 {\cal P}_- {\cal H} {\cal P}_- = J \, \bigg\{ \sum_{ X = L , R } \; \sum_{ j = 1}^{\ell - 1} 
[  S_{j , X}^+ S_{j + 1 , X}^- +  S_{j , X}^- S_{j + 1 , X}^+  + \Delta S_{j , X}^z S_{j + 1 , X}^z ] 
  \bigg\} - J_\perp \{ S_{1 , L}^+ S_{1 , R}^- + S_{1 , L}^- S_{1 , R}^+ \} + \ldots \; , 
 \label{mh.12x}
\end{equation}
\noindent
with $J_\perp \approx J_L^{'} J_{R }^{'} /(2 B)$ and the ellipses corresponding to 
subleading correction to the most relevant terms in the effective boundary Hamiltonian. 
The Hamiltonian in Eq.~(\ref{mh.12x}) again describes a single weak-link, just as $H_{\rm WL}$ in Eq.~(\ref{mh.15}).

\section{Spinless Luttinger liquid formulation of the uniform chain}
\label{sll}

Here, we review the bosonization approach to the (open boundary) XXZ-spin chain, 
which was our main theoretical tool to derive the analytical results we present in our paper.  
In doing so, we strictly follow the approach developed in Refs.~[\onlinecite{eggert92,furusaki98}],
eventually leading to the SLL-formulation of the problem \cite{hald_2}. Our reference Hamiltonian 
for an open-boundary homogeneous  XXZ-spin chain over an $\ell$-site lattice is given by 
\beq
H = J \sum_{j = 1}^{\ell - 1} \big\{ S_j^x S_{j+1}^x + S_j^y S_{j+1}^y + \Delta  S_j^z S_{j+1}^z \big\} \; . 
\label{appe.1}
\eneq
Resorting to the continuum, low-energy, long wavelength SLL description of the chain
requires introducing a spinless, real bosonic 
field $\Phi ( x   )$ and its dual field $\Theta ( x )$. The canonically 
conjugated momentum of $\Phi ( x )$ is realized, in terms of $\Theta ( x )$, as 
$ \Pi ( x ) = \frac{1}{2 \pi} \partial_x \Theta ( x )$, which implies the equal-time 
commutation relation $ [ \partial_x \Theta ( x ) , \Phi ( x' ) ] = 2 \pi i \delta ( x - x' )$ \cite{eggert92}.  
Because  throughout our paper we are interested in equal-time, equilibrium spin correlations only,
it is more useful to resort to the imaginary-time formulation for the theory of the $\Phi$ field. Letting 
$\Phi ( x , \tau )$ be the field $\Phi ( x )$ at imaginary time $\tau$, the 
corresponding imaginary time action is  given by
\beq
S_E [ \Phi ] = \frac{g}{ 4 \pi} \: \int  \: d \tau \: \int_0^L \: d x \:
\left[ \frac{1}{ u } \left( \frac{ \partial \Phi}{ \partial \tau} \right)^2
+ u \left( \frac{ \partial \Phi}{ \partial x} \right)^2 \right]
\, .
\label{appe.2}
\eneq
The parameters $g$ and $u$ in Eq.~(\ref{appe.2}) keep memory, in the effective continuum 
description of the spin chain, of the microscopic parameters in Eq.~(\ref{appe.1}). Those 
are referred to as the Luttinger parameter and the plasmon velocity, respectively, and 
are given by 
\beq
g = \frac{\pi}{2 [ \pi - {\rm arccos} (\Delta  )]} \, , \qquad
u = v_f \left[ \frac{\pi}{2} \frac{\sqrt{1 - \Delta^2}}{ 
{\rm arccos ( \Delta )}
} \right] \, , 
\label{appe.3}
\eneq
with $v_f = 2 d J$, $d$ being the lattice step (which we explicitly report here for the sake of clarity, 
though we set $d=1$ anywhere else in our paper). The ``dual'' formulation of Eq.~(\ref{appe.2}), 
involving the imaginary-time field $\Theta ( x , \tau )$, is recovered by simply substituting 
$\Phi$ with $\Theta$ and $g$ with $1/g$ \cite{giuliano05,giuliano07,giuliano09}. 
For the sake of completeness, it is worth pointing out that typically, 
within bosonization procedure, one recovers  an additional Sine-Gordon, 
Umklapp interaction term that should be added to $S_E [ \Phi ]$ in Eq.~(\ref{appe.2}).
This is better expressed as a functional of $\Theta$, and is given by \cite{giuliano05}
\beq
S^{\rm SG} [ \Theta ] = - G_U  \int_0^\beta \: d \tau \: \int_0^L \: d x \:  \cos [ 2 \sqrt{2}  \Theta ( x , \tau  ) ]  
\, . 
\label{appe.4}
\eneq
The scaling dimension of $S^{\rm SG} [ \Theta ] $ is $h_{\rm SG} = 4 g $. 
Therefore, it is always irrelevant for $1/2 < g $, while it becomes
marginally irrelevant at the ``Heisenberg point'', $g=1/2$, which we do not consider 
here and, in general,  deserves special attention and care in 
going along the bosonization procedure \cite{eggert92,sorensen}.
To account for  the open boundary conditions of
the chain, one imposes Neumann-like boundary
conditions on the field $\Phi ( x , \tau)$ at both boundaries, 
that is
\beq
\frac{ \partial \Phi ( 0 , \tau )}{ \partial x } = \frac{ \partial
\Phi ( \ell , \tau )}{ \partial x} = 0 
\, ,
\label{appe.5}
\eneq
which implies the following mode expansions
for $ \Phi ( x , \tau )$ and $\Theta ( x , \tau )$ \cite{oca,giuliano05,giuliano07,giuliano09,cmgs,gst_1}
\begin{eqnarray} 
\Phi ( x , \tau ) & = & \sqrt{\frac{2}{g}} \bigg\{ q - \frac{i \pi u \tau }{\ell} P
+ i \sum_{ n \neq 0} \frac{ \alpha ( n )}{ n} \cos \left[ \frac{ \pi n x}{\ell} 
\right] e^{ - \frac{\pi n }{\ell} u \tau} \bigg\},
\nonumber \\
\Theta ( x , \tau ) & = & \sqrt{2g} \bigg\{ \theta + \frac{ \pi x }{\ell} P
+  \sum_{ n \neq 0} \frac{ \alpha ( n )}{ n} \sin \left[ \frac{ \pi n x}{\ell} 
\right] e^{ - \frac{\pi n }{\ell} u \tau} \bigg\}
\, ,
\label{appe.6}
\end{eqnarray}
with the normal modes satisfying the algebra
\beq
[ q , P ] = i \, , \qquad [ \alpha ( n ) , \alpha (n' ) ] = n \delta_{ n + n' , 0 } \, . 
\label{appe.7}
\eneq
Finally, in terms of the continuum bosonic fields, the original spin operators are 
realized as \cite{eggert92,hikifu_1}  
\begin{eqnarray} 
S_j^+ &\longrightarrow &    \left\{ c (-1)^j
 e^{ \frac{i}{ \sqrt{2}} \Phi ( x_j , \tau ) } + b 
 e^{  
\frac{i}{ \sqrt{2}} \Phi ( x_j , \tau )  + i \sqrt{2} \Theta ( x_j , \tau )}
\right\} ,
\nonumber \\
S_j^z & \longrightarrow & \left[  \frac{1}{\sqrt{2} \pi} \frac{ \partial 
\Theta ( x_j , \tau )}{ \partial x} + a  (-1)^j 
\sin [ \sqrt{2} \Theta ( x_j ) ] \right] , 
\label{appe.8}
\end{eqnarray}
with $S_j^\pm = \frac{1}{2} [ S_j^x \pm i S_j^y]$, and 
the parameters $a,b,c$ in Eq.~(\ref{appe.8}) depending only on the 
anisotropy parameter $\Delta $.
$a,b,c$ have been numerically computed for quite a wide range of values of 
the system parameters. Since they are actually not essential to the analytic RG
analysis (which is the only reason why we have to consider the Luttinger liquid formulation 
of the spin chain), we refer the interested reader to the literature \cite{hikifu_1,hikifu_2,shashi,lky_1,lky_2}. 
Using the bosonization formalism, it is possible to generalize to a finite imaginary 
time difference $\tau $ the spin-spin correlation functions for the chain with 
open boundary conditions at its endpoints, this generalizing the 
results for the equal-time  spin-spin correlation functions, derived in 
Ref.~[\onlinecite{hikifu_1}] and extended in Ref.~[\onlinecite{hikifu_2}] to the case of a 
nonzero uniform magnetic field in the chain. The derivation of the finite-$\tau $ correlation functions
$G_{+ , -}  ( x , x'  ; \tau | \ell ) = 
\langle {\bf T}_\tau S^+_x ( \tau ) S_{x'}^- ( 0 ) \rangle $ and 
$G_{z , z}  ( x , x'  ; \tau | \ell ) = 
\langle {\bf T}_\tau S^z_x ( \tau ) S_{x'}^z ( 0 ) \rangle $,
is discussed in detail in Ref.~[\onlinecite{gst}]. Here, we just quote the main result, which we have diffusely used 
in the body of our paper. One obtains 
\begin{eqnarray}
&& G^{+-} ( x , x' ; \tau  | \ell ) = \nonumber \\
&& c^2 (-1)^{ x - x'} \left| \alpha(x) \right|^{\frac{1}{4 g }} \left| \alpha(x') \right|^{\frac{1}{4 g }} 
\left| \frac{2\ell}{ \pi} \sinh \left[ \zeta_\tau   \right] \right|^{ - \frac{1}{2g}} 
\left| \frac{2\ell}{\pi} \sinh \left[ \zeta_\tau(x+x') \right] \right|^{ - \frac{1}{2g}} 
\nonumber \\
&& + b^2 \left| \alpha(x) 
\right|^{\frac{1}{4 g }-g} \left| \alpha(x') \right|^{\frac{1}{4 g }-g} 
 \left| \frac{2\ell}{\pi} \sinh \left[ \zeta_\tau (x-x') \right] \right|^{ - \frac{1}{2g}-2g} 
\left| \frac{2\ell}{\pi} \sinh \left[ \zeta_\tau(x+x') \right] \right|^{ - \frac{1}{2g}+2g} 
\nonumber \\
&& + bc \;  {\rm sgn} ( x - x' ) \left| \alpha(x) \right|^{\frac{1}{4 g }} \left| \alpha(x') \right|^{\frac{1}{4 g }} 
\left| \frac{2\ell}{ \pi} \sinh \left[ \zeta_\tau(x-x') \right] \right|^{ - \frac{1}{2g}} 
\left| \frac{2\ell}{ \pi} \sinh \left[ \zeta_\tau (x+x') \right] \right|^{ - \frac{1}{2g}} 
\left[ \frac{(-1)^x}{\left| \alpha(x') \right|^{g}} - \frac{(-1)^{x'}}{\left| \alpha(x) \right|^{g}} \right] ,
\label{corr.x1}
\end{eqnarray}
as well as 
\begin{eqnarray}
&& G^{zz} ( x , x' ; \tau | \ell  )  = 
- \frac{g}{4 \ell^2} \biggl\{ \frac{1 - \cosh \big[ \pi u  \tau /\ell \big] \; \cos \big[ \pi  ( x - x' )/\ell \big]}
{1 + \cos^2 \big[ \pi  ( x - x' ) / \ell \big] - 2 \cos \big[ \pi ( x - x') / \ell \big] 
\cosh \big[ \pi u \tau/\ell \big] + \sinh^2 \big[ \pi u \tau / \ell \big] } \nonumber \\
&& \hspace*{3.4cm} + \frac{1 - \cosh \big[ \pi u \tau / \ell \big] \cos \big[ \pi (x + x') / \ell \big] }
{1 + \cos^2 \big[ \pi ( x + x' ) / \ell \big] - 2 \cos \big[ \pi ( x + x') / \ell \big]
\cosh \big[ \pi u \tau / \ell \big] + \sinh^2 \big[ \pi u \tau / \ell \big] } \biggr\}
\nonumber \\
&& + \frac{a^2}{2} (-1)^{x- x'}  
\left| \alpha(x) \right|^{-g} \left| \alpha(x') \right|^{-g}  \biggl\{ 
\left| \frac{ \sinh \left[ \zeta_\tau (x-x') \right]}{ \sinh \left[ \zeta_\tau(x+x') \right] } \right|^{-2g} 
- \left| \frac{ \sinh \left[ \zeta_\tau (x-x') \right]}{ \sinh \left[ \zeta_\tau(x+x') \right] } \right|^{2g} 
\biggr\}
\nonumber \\
&& - \frac{a ig}{2 \ell} (-1)^{x'} \left| \alpha(x') \right|^{-g} 
\biggl\{ \coth \left[ \zeta_\tau(x+x') \right] - \coth \left[ \zeta_\tau(-x-x') \right] 
 - \coth \left[ \zeta_\tau(x-x') \right] + \coth \left[ \zeta_\tau(x'-x) \right]  \biggr\}
\nonumber \\
&& - \frac{a ig}{2 \ell} (-1)^x
\left| \alpha(x) \right|^{-g} \biggl\{ \coth \left[ \zeta_\tau(x+x') \right] - \coth \left[ \zeta_\tau(-x-x') \right] 
+ \coth \left[ \zeta_\tau(x-x')  \right] - \coth \left[ \zeta_\tau(x'-x) \right]  \biggr\} \; ,
\label{corr.x2}
\end{eqnarray}
where we defined $\alpha(x) = \frac{2 \ell}{\pi} \sin(\pi x/ \ell)$,
and also $\zeta_\tau(x) = \frac{\pi}{2\ell} [ u  \tau + i x ]$
The correlation functions in Eqs.~(\ref{corr.x1}, \ref{corr.x2}) are the main ingredient we used throughout 
our paper to analytically discuss the properties of our system.

\section{Renormalization group flow of the running coupling strengths}
\label{reng}

We here review the derivation of the RG equations for the running couplings 
associated to the effective Kondo Hamiltonian $H_K$. To do so, we employ the framework used 
in Ref.~[\onlinecite{gst}], eventually generalized to the case of a nonzero 
$B$ applied to ${\bf S}_{\bf G}$. To recover the RG equations for the boundary running coupling 
associated to $H_K$, we resort to the imaginary-time SLL formalism of appendix~\ref{sll}. 
The weak-coupling assumption for the boundary couplings, $J'_{L (R) } / J < 1$, $J'_{ z , L (R)} / J < 1$, 
allows us to separately bosonize the two leads, which eventually allows us for trading $H_K$ for the Kondo action $S_K$ given by 
\begin{eqnarray}
S_K &=& \int \: d \tau \: \Biggl\{ [ J_L^{'}  e^{ \frac{i}{\sqrt{2}} \Phi_L ( 0 , \tau) } + J_R^{'}  e^{ \frac{i}{\sqrt{2}} \Phi_R (0 ,  
\tau) } ] S_{\bf G}^- ( \tau ) + 
[ J_L^{'}  e^{ - \frac{i}{\sqrt{2}} \Phi_L ( 0 , \tau) } + J_R^{'}  e^{- \frac{i}{\sqrt{2}} \Phi_R (0 ,  \tau) } ] S_{\bf G}^+ ( \tau )  \nonumber \\
&& \hspace*{1.2cm} +\left[ \frac{J_{z , L}^{'} }{\sqrt{2} \pi} \frac{\partial \Theta_L ( 0 , \tau)}{\partial x} + \frac{J_{z , R}^{'} }{\sqrt{2} \pi} 
\frac{\partial \Theta_R ( 0 , \tau)}{\partial x} \right] S_{\bf G}^z ( \tau ) + B S_{\bf G}^z ( \tau)  \Biggr\} \; . 
\label{rengu.1}
\end{eqnarray}
\noindent
As we are interested in the behavior of the system as its size grows, we use $\ell$ as the running scale 
of the RG flow. Introducing a scale $\ell_0$ (of the order of the lattice step) 
as a reference length, we define the running 
boundary couplings $G_{L (R)} ( \ell ) , G_{z , L (R)} ( \ell )$ as \cite{gst} 
\begin{eqnarray}
G_{L (R)}  ( \ell )  &=&  \frac{1}{2} \left( \frac{\ell}{\ell_0} \right)^{1 - \frac{1}{2g} } \: \frac{J'_{ L ( R ) }}{J}, \nonumber \\
G_{z , L (R)}  ( \ell ) &=& \frac{1}{2} \: \frac{J'_{z ,  L ( R ) }}{J}
\, . 
\label{rengu.01}
\end{eqnarray}
\noindent
To derive the RG equations for the running couplings, we employ a boundary 
version of the technique based on the operator product expansion (OPE) discussed by Cardy within the 
context of deformed conformal field theories \cite{cardy_1}. In principle, higher order OPEs can
induce contributions  mixing the $_L$ and $_R$ boundary couplings, such as terms 
$\propto \cos \left[ \frac{1}{ \sqrt{2}} ( \phi_L ( 0 ) - \phi_R ( 0 ) ) \right]$. Such terms correspond to 
channel-mixing contributions to the boundary Hamiltonian. In the context of two-channel electronic Kondo 
effect, they would imply that only a single, ``hybridized'', electronic channel couples to the magnetic impurity, thus 
switching back to single-channel Kondo effect. Here, since those terms are bilinear
operators of the spin densities at the two sides of the impurity, they correspond to irrelevant or, more generally,
subleading correction to the boundary Hamiltonian \cite{gst}. Accordingly, neglecting them and assuming $B / J \ll 1$, 
we obtain 
\begin{eqnarray}
 \frac{d G_{L (R)} ( \ell) }{d \ln \left( \frac{\ell}{\ell_0} \right)} &=& \left( 1 - \frac{1}{2 g} \right) G_{L (R)} ( \ell) + 
   \cosh \left( \frac{B \ell}{2 J} \right)  \:  G_{L (R)} ( \ell )   G_{z , L (R)} ( \ell ), \nonumber \\
  \frac{d G_{z , L (R)} ( \ell) }{d \ln \left( \frac{\ell}{\ell_0} \right)} &=&  
 \cosh \left( \frac{B \ell}{  J} \right) \:  [ G_{L (R)}( \ell ) ]^2
 \, . 
 \label{rengu.2}
\end{eqnarray}
\noindent
As $B \to 0$, Eqs.~(\ref{rengu.2}) reduce back to the ones implemented 
in Ref.~[\onlinecite{gst}] to derive the Kondo screening 
length in various regimes. In that case, it is possible to provide a closed-form solution 
for the integral curves, from which one can extract the analytical expression of $\xi_K$. 
For the sake of completeness,   we now review the analytical 
derivation of $\xi_K$ for $B=0$. In this case, one may simplify 
Eqs.~(\ref{rengu.2}) by introducing the boundary coupling strengths $X_{L ( R )} ( \ell )= G_{L (R)} ( \ell)$ and   
$X_{z , L (R)} ( \ell )=  1 - \frac{1}{2 g} + G_{z , L (R)} ( \ell ) $.
In terms of the novel running coupling strengths, Eqs.~(\ref{rengu.2}) take the form 
\begin{eqnarray}
 && \frac{ d X_X ( \ell ) }{ d \ln ( \frac{\ell}{\ell_0} )} =
X_X (\ell ) X_{z , X} ( \ell ) \nonumber \\
&& \frac{ d X_{z , X} ( \ell ) }{ d \ln ( \frac{\ell}{\ell_0} )} =
X_X^2 ( \ell )
\, , 
\label{mh.6}
\end{eqnarray}
with $X = L , R$. To integrate Eqs.~(\ref{mh.6}), we note that there is 
an RG-invariant  $\kappa_{X}$, given 
\beq
\kappa_{X}  = (X_{z , X}  ( \ell ))^2 - (X_{X}  ( \ell ))^2 \, . 
\label{mh.8}
\eneq
From the integral curves of Eqs.~(\ref{mh.6}), one estimates $\xi_K$ as the 
length scale at which the  boundary couplings enter the nonperturbative regime. 
As Eqs.~(\ref{mh.6}) are separated in the $_{L-R}$-indices, 
in the following we drop the corresponding labels in the running couplings.
$\xi_K$ is estimated as the scale at which the running couplings diverge and, 
clearly, its  functional form explicitly depends  on the sign of $\kappa$ \cite{gst}. Specifically,
one obtains:
\begin{itemize}
 \item {\bf $\kappa=0$}. In this case $X ( \ell ) = X_z ( \ell)$, with 
the explicit solution given by  
\beq
X_z (\ell) = \frac{X_z (\ell_0 )}{ 1 - X_z ( \ell_0 ) \, \ln ( \ell / \ell_0 ) } \, . 
\label{mh.9}
\eneq
\noindent
From Eq.~(\ref{mh.9}), one obtains 
\begin{equation}
\xi_K  \sim \ell_0 \exp \left[  \frac{1}{X_z (\ell_0 ) } \right] \, ,
\label{mh.10}
\end{equation}
\noindent
which is the familiar result one recovers for the ``standard'' Kondo
effect in metals \cite{affleck_length}.
\item {\bf  $\kappa<0$}. In this case, the explicit solution of Eqs.~(\ref{mh.6}) is given by 
\begin{equation}
 X_z ( \ell ) = \sqrt{- \kappa } \tan \left\{ {\rm atan} \left[ \frac{X_z ( \ell_0 ) }{\sqrt{- \kappa}} \right] + 
 \sqrt{- \kappa} \, \ln \left( \frac{\ell}{\ell_0} \right) \right\} \,, \qquad
 X ( \ell ) = \sqrt{-\kappa + X_z^2 ( \ell ) } \; , 
 \label{mh.11}
\end{equation}
which yields 
\beq
\xi_K \sim \ell_0 \exp \left[ \frac{ \pi  - 2 \:
 {\rm atan} ( X_z (\ell_0) /  \sqrt{ | \kappa | } ) }
{2 \sqrt{ | \kappa | }} \right] \; . 
\label{mh.12}
\eneq
\noindent

\item {\bf $\kappa > 0$}. In this case 
\begin{equation}
 X_z ( \ell ) = - \sqrt{\kappa} \left\{ \frac{[ X_z ( \ell_0 )  - \sqrt{\kappa}] \; \left( \ell / \ell_0 \right)^{2\sqrt{\kappa}} + [ X_z ( \ell_0 ) + \sqrt{\kappa} ]  }{
 [ X_z ( \ell_0 )  - \sqrt{\kappa} ] \; \left( \ell / \ell_0 \right)^{2\sqrt{\kappa}} - [ X_z ( \ell_0 ) + \sqrt{\kappa} ] } \right\} \,, \qquad
 X ( \ell ) = \sqrt{-\kappa + X_z^2 ( \ell ) } \; .
 \label{mh.13}
\end{equation}
\noindent
As a result, we obtain 
\beq
\xi_K \sim \ell_0 \left\{    \frac{X_z (\ell_0 ) + \sqrt{\kappa}}{
X_z (\ell_0 ) - \sqrt{\kappa} }  \right\}^\frac{1}{2 \sqrt{\kappa}} \, . 
\label{mh.14}
\eneq
\noindent
\end{itemize}
Remarkably, it is also possible to recast Eqs.~(\ref{mh.9}, \ref{mh.11}, \ref{mh.13}) into a scaling form, 
in which $\ell_0$ is traded for an explicit dependence on $\xi_K$ of the running couplings which, accordingly, 
become a function of the dimensionless running parameter $\ell / \xi_K$. In particular, for $\ell  < \xi_K$, 
one obtains 
\begin{equation}
 X_z (\ell) = \left\{ \begin{array}{ll} 
 \displaystyle \frac{1}{ \ln \left( \frac{\ell}{\xi_K} \right) } & (\kappa = 0 ), \\
 \displaystyle \sqrt{- \kappa } \tan \left\{ \frac{\pi}{2} + \sqrt{- \kappa} \ln \left( \frac{\ell}{\xi_K} \right)   \right\} & (\kappa < 0 ), \\
 \displaystyle \sqrt{\kappa} \left\{ \frac{ \left( \xi_K / \ell \right)^{\sqrt{2 \kappa}} + 1  }{ \left( \xi_K / \ell \right)^{\sqrt{2 \kappa}} - 1 } \right\}
 & (\kappa > 0 ) \;. \end{array} \right.
 \label{mhscle}
\end{equation}
In general, no closed-form solutions can be recovered for $B \neq 0$, which therefore implies 
numerically solving Eqs.~(\ref{rengu.2}), as we did to  discuss the finite-$B$ case. 

\bibliography{kondo_xxz.bib}

\end{document}